\documentclass[preprint, linenumber]{aastex61}

\usepackage{amsmath}
\usepackage{graphicx}
\usepackage{natbib}
\usepackage{ulem}
\usepackage{bm}

\usepackage{xspace}
\usepackage{hyperref}

\usepackage{mathtools}
\usepackage{stmaryrd}
\usepackage{multirow}

\defcitealias{2007PhPl...14e2101A}{AG07}
\defcitealias{1996ApJ...472..398B}{BDBG96}
\defcitealias{2011ApJ...728..147C}{C11}
\defcitealias{1989CourantHilbert}{CH89}
\defcitealias{1982SoPh...76..239E}{ER82}
\defcitealias{1983SoPh...88..179E}{ER83}
\defcitealias{1986NASCP2449..347E}{ER86}
\defcitealias{2011SoPh..272..119F}{F11}
\defcitealias{2011SSRv..158..289G}{GER11}
\defcitealias{2013ApJ...766...55K}{K13}
\defcitealias{2015ApJ...801...23L}{LN15b}
\defcitealias{2015ApJ...810...87L}{LN15a}
\defcitealias{2014ApJ...789...48O}{ORT14}
\defcitealias{2015ApJ...806...56O}{ORT15}
\defcitealias{1974Whitham}{W74}
\defcitealias{2021SoPh..296...95Y}{Y21}

\renewcommand{\vec}[1]{\ensuremath{\bm{#1}}}

\newcommand{\uvec}[1]{\ensuremath{\bm{e}_{#1}}}
\newcommand{\Exp}[1]{\ensuremath{{\rm e}^{#1}}}

\newcommand{\Alf}{Alfv$\acute{\rm e}$n}
\newcommand{\Alfnic}{Alfv$\acute{\rm e}$nic}

\newcommand{\mathd}{\ensuremath{{\rm d}}}

\newcommand{\kappai}{\ensuremath{\kappa_{\rm i}}}

\newcommand{\mi}{\ensuremath{m_{\rm i}}}

\newcommand{\ptot}{\ensuremath{p_{\rm tot}}}

\newcommand{\rhoi}{\ensuremath{\rho_{\rm i}}}

\newcommand{\Rm}{\ensuremath{R_{\rm m}}}

\newcommand{\va}{\ensuremath{v_{\rm A}}}
\newcommand{\vai}{\ensuremath{v_{\rm Ai}}}

\newcommand{\vph}{\ensuremath{v_{\rm ph}}}

\newcommand{\ckL}{\ensuremath{c_{\rm kL}}}
\newcommand{\xint}{\ensuremath{x_{\rm i}}}
\newcommand{\xext}{\ensuremath{x_{\rm e}}}
\newcommand{\rhoL}{\ensuremath{\rho_{\rm L}}}
\newcommand{\rhoR}{\ensuremath{\rho_{\rm R}}}
\newcommand{\vaL}{\ensuremath{v_{\rm AL}}}
\newcommand{\vaR}{\ensuremath{v_{\rm AR}}}

\newcommand{\mL}{\ensuremath{m_{\rm L}}}
\newcommand{\mR}{\ensuremath{m_{\rm R}}}

\newcommand{\kappaL}{\ensuremath{\kappa_{\rm L}}}
\newcommand{\kappaR}{\ensuremath{\kappa_{\rm R}}}

\newcommand{\sres}{\ensuremath{s_{\rm res}}}

\newcommand{\vgy}{\ensuremath{v_{{\rm gr},y}}}
\newcommand{\vgz}{\ensuremath{v_{{\rm gr},z}}}
\newcommand{\vpy}{\ensuremath{v_{{\rm ph},y}}}

\shorttitle{Oblique Quasi-Kink Modes in Asymmetric Coronal Slabs}
\shortauthors{Chen et al.}
\begin{document}

\title{Oblique Quasi-Kink Modes in Solar Coronal Slabs Embedded in an Asymmetric Magnetic Environment: Resonant Damping, Phase and Group Diagrams}

\correspondingauthor{Bo Li}
\email{bbl@sdu.edu.cn}

\author{Shao-Xia Chen}
\affiliation{Shandong Provincial Key Laboratory of Optical Astronomy and Solar-Terrestrial Environment,
   Institute of Space Sciences, Shandong University, Weihai 264209, China}

\author{Bo Li}
\affiliation{Shandong Provincial Key Laboratory of Optical Astronomy and Solar-Terrestrial Environment,
   Institute of Space Sciences, Shandong University, Weihai 264209, China}

\author{Mingzhe Guo}
\affiliation{Shandong Provincial Key Laboratory of Optical Astronomy and Solar-Terrestrial Environment,
   Institute of Space Sciences, Shandong University, Weihai 264209, China}

\author{Mijie Shi}
\affiliation{Shandong Provincial Key Laboratory of Optical Astronomy and Solar-Terrestrial Environment,
   Institute of Space Sciences, Shandong University, Weihai 264209, China}

\author{Hui Yu}
\affiliation{Shandong Provincial Key Laboratory of Optical Astronomy and Solar-Terrestrial Environment,
   Institute of Space Sciences, Shandong University, Weihai 264209, China}

\begin{abstract}
There has been considerable interest in magnetoacoustic waves in
    static, straight, field-aligned, one-dimensional equilibria where the exteriors of a magnetic slab are different between the two sides.
We focus on trapped, transverse fundamental, oblique quasi-kink modes
    in pressureless setups where the density
    varies continuously from a uniform interior
    (with density $\rho_{\rm i}$) to a uniform
    exterior on either side (with density $\rho_{\rm L}$ or $\rho_{\rm R}$),
    assuming $\rho_{\rm L}\le\rho_{\rm R}\le\rho_{\rm i}$. 
The continuous structuring and oblique propagation make our study new relative
    to pertinent studies, and lead to wave damping
    via the Alfv$\acute{\rm e}$n resonance.
We compute resonantly damped quasi-kink modes as resistive eigenmodes, 
    and isolate the effects of system asymmetry by varying
    $\rho_{\rm i}/\rho_{\rm R}$ from the ``Fully Symmetric''
    ($\rho_{\rm i}/\rho_{\rm R}=\rho_{\rm i}/\rho_{\rm L}$) 
    to the ``Fully Asymmetric'' limit ($\rho_{\rm i}/\rho_{\rm R}=1$).   
We find that the damping rates possess a nonmonotonic
    $\rho_{\rm i}/\rho_{\rm R}$-dependence as a result of the difference between
    the two Alfv$\acute{\rm e}$n continua, and resonant absorption occurs
    only in one continuum when $\rho_{\rm i}/\rho_{\rm R}$ is below some threshold.
We also find that the system asymmetry results in two qualitatively different
    regimes for the phase and group diagrams.
The phase and group trajectories lie essentially on the same side
    (different sides) relative to the equilibrium magnetic field
    when the configuration is not far from
    a ``Fully Asymmetric'' (``Fully Symmetric'') one. 
Our numerical results are understood by making analytical progress
    in the thin-boundary limit, and discussed for imaging observations
    of axial standing modes and impulsively excited wavetrains.
\end{abstract}

\keywords{magnetohydrodynamics (MHD) --- Sun: corona --- Sun: magnetic fields  --- waves}

\section{INTRODUCTION}
\label{sec_intro}
It is well accepted that the highly structured solar atmosphere 
   hosts a rich variety of low-frequency
   magnetohydrodynamic (MHD) waves and oscillations
   \citep[see e.g.,][for reviews]{2015SSRv..190..103J,2015LRSP...12....6K,2020SSRv..216..136L,2021SSRv..217...34W,2021SSRv..217...76B}.
When observed, these waves/oscillations tend to be placed in the context
   of either atmospheric heating \citep[see the reviews by e.g.,][]{2015RSPTA.37340269D,2015RSPTA.37340261A,2020SSRv..216..140V}    
   or solar atmospheric seismology 
   \citep[SAS, for reviews, see e.g.,][]{2005LRSP....2....3N,2012RSPTA.370.3193D,2020ARA&A..58..441N}.
Whichever the context, a thorough theoretical understanding
   on MHD waves in structured media proves indispensable given the need to, say,
   pinpoint the physical identity of an observed oscillatory signal in the first place.
Consequently, extensive use has long been made of equilibrium configurations
   where the physical parameters are structured 
   only in one transverse direction, 
   in both cylindrical
   (e.g., \citealt{1970A&A.....9..159R,1975IGAFS..37....3Z, 1979A&A....76...20W,1983SoPh...88..179E}, hereafter~\citetalias{1983SoPh...88..179E})
   and planar geometries 
   (e.g., 
   \citealt{1978ApJ...226..650I,1979ApJ...227..319W,1981SoPh...69...27R,1981SoPh...69...39R,1982SoPh...76..239E}, \citetalias{1982SoPh...76..239E} hereafter). 
While ``a first approximation of reality''
    \citep[][p.446]{2006RSPTA.364..433G},       
   one-dimensional~(1D) equilibria remain in routine use
   given the~(semi-)analyitcal treatments they permit
   and/or the relevant wave physics they help elucidate.

Much progress has been made in the past two decades for 
   cylindrical implementations of 1D equilibria. 
Let ``\citetalias{1983SoPh...88..179E} equilibria'' refer
   to the canonical, straight, field-aligned, static 
   configurations addressed by \citetalias{1983SoPh...88..179E}.
Let ``\citetalias{1983SoPh...88..179E}-like equilibria''
   refer further to those that differ from
   the \citetalias{1983SoPh...88..179E} equilibria only
   by replacing the step transverse profiles therein
   with continuous ones. 
An extensive set of studies then indicated
   that the \citetalias{1983SoPh...88..179E} and/or \citetalias{1983SoPh...88..179E}-like equilibria still yield new physics, 
   to illustrate which point we name only a few examples. 
To start, revisiting kink modes in an \citetalias{1983SoPh...88..179E} equilibrium
   has enabled one to better understand
   both their physical nature \citep[e.g.,][]{2009A&A...503..213G,2012ApJ...753..111G,2014ApJ...788....9G}
   and their energy-carrying capabilities
   \citep[e.g.,][]{2013ApJ...768..191G,2014ApJ...795...18V}.
Likewise, recent examinations on coronal sausage modes 
   in either an \citetalias{1983SoPh...88..179E} \citep{2014ApJ...781...92V}
   or an \citetalias{1983SoPh...88..179E}-like setup \citep[e.g.,][]{2015ApJ...810...87L,2017ApJ...836....1Y}
   have shed new light on the wave behavior in the neighborhood of
   the critical axial wavenumbers that separate the trapped from
   the leaky regime. 
Furthermore, destructive interference has gained new attention
   \citep[see][and references therein for motivating ideas]{1991JPlPh..45..453C}
   as a unifying process that underlies
   the key notions of 
       lateral leakage
       \citep[e.g.,][]{2007PhPl...14e2101A,2015ApJ...806...56O,2022ApJ...928...33L},
   resonant damping, and phase mixing 
        (\citealt{2002ApJ...577..475R,2015ApJ...803...43S}; also references therein).	
Note that these notions themselves are 
   not specific to a particular mode, an example being 
   the relevance of resonant damping to both kink and sausage modes
   \citep[e.g.,][]{2016ApJ...823...71G,2021A&A...646A..86G,2021ApJ...908..230C}.
Note also that new physics has also been gathered by considering those equilibria
   where the inhomogeneity can be rendered 1D by appropriate coordinate transformations
   but the configuration itself may differ considerably from \citetalias{1983SoPh...88..179E}.
The adoption of elliptic coordinates, for instance, 
   yields a clear distinction between
   differently polarized kink modes in coronal loops
    with elliptic cross-sections 
   (e.g., \citealt{2003A&A...409..287R,2009A&A...494..295E}; also \citealt{2011A&A...527A..53M,2020ApJ...904..116G}).
Likewise, the application of bicylindrical coordinates to a system of
   two parallel loops enables one to address how a classic kink mode 
   in an isolated loop splits into different kink-like oscillations that are polarized 
   differently with respect to the orientation of the system
   (\citealt{2008A&A...485..849V}; also \citealt{2008ApJ...676..717L,
   2009ApJ...692.1582L,2010A&A...515A..33R}),
   and how the resonant damping of these kink-like oscillations is affected
   by, say, the separation between the two loops
   (\citealt{2011A&A...525A...4R,2014A&A...562A..38G}; see also 
   \citealt{2015A&A...582A.120S}).

Planar implementations of 1D equilibria have also proven to be fruitful.
Let an ``\citetalias{1982SoPh...76..239E} equilibrium'' refer
   to the canonical slab configuration examined in \citetalias{1982SoPh...76..239E},
   and restrict ourselves to only two groups of studies where
   \citetalias{1982SoPh...76..239E} is taken as a prototype.
The first group focuses on curved slabs, motivated 
   either by vertically polarized kink modes in active region (AR) loops
   first imaged by TRACE \citep[e.g.,][]{2004A&A...421L..33W,2008A&A...489.1307W} 
   or by the TRACE \citep[e.g.,][]{2002SoPh..206...69S,2004SoPh..223...77V} 
      and SDO/AIA observations \citep[e.g.,][]{2015ApJ...804L..19J,2019ApJ...880....3A} 
   that the response of coronal arcades to neighboring eruptions 
      may involve the entire arcade rather than only individual structures embedded therein. 
Let $\vec{B}_0$ denote the equilibrium magnetic field.
Let $\phi$ be the coordinate along $\vec{B}_0$, and $r$ 
    a transverse coordinate.
The equilibria are 1D in that the equilibrium quantities
    depend only on $r$, the third ($y$-) direction being ignorable.  
Some new insights then arise for, say, fast modes 
    even when the $y$-propagation is prohibited 
    (the $y$-wavenumber $k_y=0$).
As examples, it was found that the $r$-slopes of the equilibrium quantities
    are crucial in determining whether fast modes can be trapped
    \citep[e.g.,][]{2006A&A...446.1139V,2016A&A...593A..52P}, 
    and wave leakage into the ambient may need to surmount an evanescent barrier
    \citep[e.g.,][]{2005A&A...438..733B,2006A&A...449..769V}.
If a non-vanishing $k_y$ is further considered, then fast modes were shown to
    possess mixed polarizations in that their velocity perturbations 
    involve the components both in and out of the $r-\phi$ plane
    (\citealt{2017A&A...608A.108T,2022MNRAS.514.4329L}; also \citealt{2010ApJ...713..651R,2015ApJ...814..105H}).
Additional insights were also obtained in connection with the
    \Alf\ continuum, two notable examples being that
    fast wave energy may be transferred to \Alfnic\ motions in the ambient 
    \citep{2013ApJ...763...16R}
    and that a new fast mode, heavily damped spatially, may occur
    when one sees the frequency rather than wavenumber as real-valued
    in the relevant eigenvalue problem (EVP) \citep{2018ApJ...858....6H}. 
Note that the clear distinction between kink and sausage modes
    in \citetalias{1982SoPh...76..239E}
    tends not to hold \citep[e.g.,][Figure~4]{2006A&A...455..709D}, 
    the reason largely being that the equilibrium quantities in the ambient
    are not symmetric about the curved slab.
Evidently, this imperfect distinction is not specific to curved configurations,
    and has in fact been the focus of the second group of recent studies.
    
The 1D equilibria addressed in the second group are
    not far from \citetalias{1982SoPh...76..239E}.    
Let $(x, y, z)$ denote a Cartesian coordinate system, and 
    let the equilibrium magnetic field~$\vec{B}_0$ be aligned with the $z$-axis. 
The 1D equilibria are now structured only in $x$, being 
    invariant and infinitely extended in $y$. 
As in \citetalias{1982SoPh...76..239E}, three uniform regions
    are discriminated, the internal one being a slab 
    and the other two being its exteriors
    \footnote{See \citealt{2018ApJ...868..128S,
    2019FrASS...6...48A} where an arbitrary number of layers are allowed.}. 
Different from \citetalias{1982SoPh...76..239E}, however, is that 
    the environment is asymmetric, namely  
    the equilibrium quantities in one exterior are different from those 
    in the other. 
A considerable number of EVP studies were devoted to
    magnetoacoustic waves propagating in the $x-z$ plane, 
    with the initial efforts addressing
           nonmagnetic \citep[e.g.,][]{2017SoPh..292...35A, 2018ApJ...855...90A}
    versus magnetic exteriors \citep[e.g.,][]{2018ApJ...853..136K}. 
Further addressed are such effects as time-stationary flows in
    the interior \citep[e.g.,][]{2018SoPh..293...86B,2022ApJ...935...41Z}
    or exterior \citep{2022ApJ...937...23Z}, 
    and the construction of axial standing modes with propagating ones
    \citep[e.g.,][]{2020ApJ...890..109O,2020ApJ...898...19O}.  
While sometimes rather complicated, this series of 1D equilibria turns out to be
    tractable semi-analytically.
It is just that in general the resulting dispersion relations (DRs) do not factorize into
    independent expressions that govern kink and sausage modes individually. 
Nonetheless, the spatial behavior of the eigenfunctions still allows such terms
    as quasi-kink and quasi-sausage modes to be proposed
    \citep[e.g.,][Figure~3]{2017SoPh..292...35A}.       
Overall, this series of studies demonstrated that the differences
    from \citetalias{1982SoPh...76..239E}
    in terms of the dispersion properties
    tend to be observationally relevant for, say,
    magnetic bright points \citep{2018ApJ...868..128S}, 
    light bridges \citep{2021ApJ...906..122Z},
    and flanks of coronal mass ejections 
       (CMEs, \citealt{2018SoPh..293...86B}).  
Conversely, the seismological techniques
    based on, say, amplitude ratios and/or minimum perturbation shifts
    can be invoked to infer how significantly one exterior 
    differs from the other, a proposal that was both
    conceptually outlined \citep[e.g.,][]{2018ApJ...855...90A,2022ApJ...934..155Z}
    and observationally applied \citep[e.g.,][]{2018SoPh..293...86B,2019FrASS...6...48A}.
      
This study is intended to present an EVP study on
    trapped, oblique, quasi-kink modes in
    straight, field-aligned, coronal slabs 
    embedded in an asymmetric environment, the focus being on
    how the relevant dispersion properties are affected by the differences
    in one exterior from the other.  
Zero-beta MHD will be adopted, to comply with which
    a uniform equilibrium magnetic field $\vec{B}_0$ is taken.
The structuring is therefore solely in the equilibrium density $\rho_0(x)$,
    from which a uniform interior (with density $\rhoi$)
    and two uniform exteriors (with densities $\rhoL$ and $\rhoR$)
    are identified.
Note that the exteriors will be referred to as ``left'' and ``right'' for the
    ease of description, and hence the subscripts~${\rm L}$ and~${\rm R}$.     
By so doing, the asymmetry of the system is entirely encapsulated 
    in the density asymmetry, namely the difference between $\rhoL$ and $\rhoR$.
Our study is new in the following aspects.
Firstly, we will simultaneously incorporate a continuous $\rho_0(x)$
    and a non-vanishing out-of-plane wavenumber $k_y$, 
    making it inevitable for quasi-kink modes to be damped
    by the \Alf\ resonance. 
The need to incorporate the two factors can be seen as natural.  
The density asymmetry, however, means that two regimes may be distinguished
    by whether the resonance occurs in only one \Alf\ continuum or in both continua.
This distinction, in turn, may have observational implications for, say,
    vertically polarized kink modes. 
Secondly, we will address how the density asymmetry affects
    the phase and group diagrams of oblique quasi-kink modes.
This aspect of dispersion properties has not been examined in the literature
    to our knowledge, but is of general importance 
    for understanding the large-time behavior
    of the system in response to impulsive and localized exciters.       
As such, our theoretical examination is expected to be
    relevant for a rather broad range of wave observations
    in slab-like structures, two examples being Sunward-moving dark tadpoles
    in post-flare supra-arcades \citep{2005A&A...430L..65V}, 
    and CME-induced cyclic transverse motions of streamer stalks
    (streamer waves; e.g., \citealt{2010ApJ...714..644C,2013ApJ...766...55K,2020ApJ...893...78D}).

This manuscript is structured as follows.
Section~\ref{sec_ProbFormu} formulates our EVP and details 
    its numerical solution procedure. 
The thin-boundary limit is examined in 
    Section~\ref{sec_TB}, where we address the EVP (semi-)analytically
    such that our numerical results can be validated and better understood.
Section~\ref{sec_ResIRA} then focuses on the resonant damping of quasi-kink modes,
    with our results on the phase and group diagrams collected in
    Section~\ref{sec_ResIIgrp}. 
We summarize this study in Section~\ref{sec_conc}.

\section{Problem Formulation}
\label{sec_ProbFormu}
  
\subsection{Equilibrium and Overall Description}
\label{sec_ProbFormu_sub_EQ}

\begin{figure}
\centering
\includegraphics[width=.85\columnwidth]{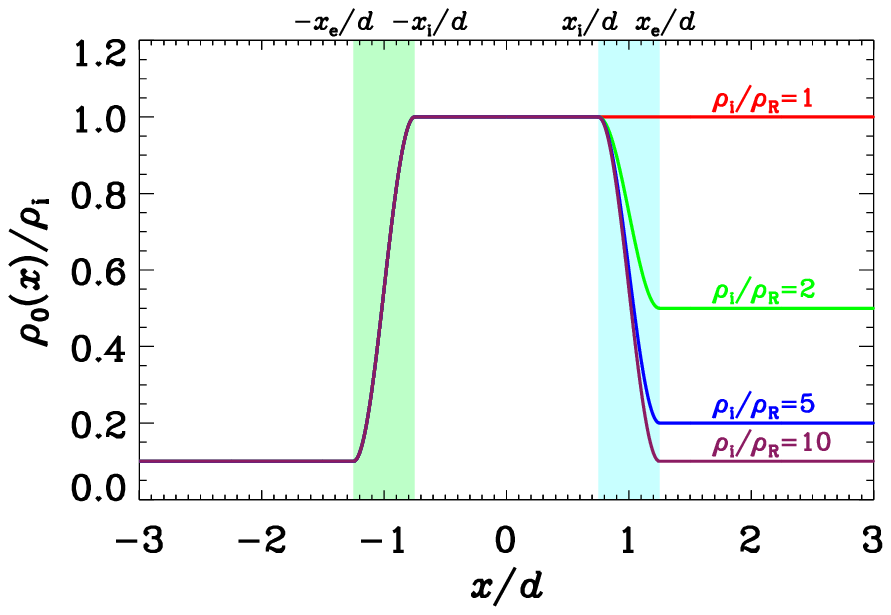}
\caption{
Transverse density profile $\rho_0(x)$ 
   for the examined asymmetric slab system. 
Two transition layers (TLs, the shaded portions), centered around the nominal 
   slab boundaries ($|x|=d$) and both of width~$l$,
   are placed symmetrically about the nominal slab axis ($x=0$).
Some smooth profile is adopted 
   to connect the interior density $\rhoi$
   to $\rhoL$ ($\rhoR$), the density in the left 
   (right) exterior (see Equation~\eqref{eq_rhoEQ}).
A pair $[\rhoi/\rhoL, l/d] = [10, 0.5]$ is adopted,
   while a number of $\rhoi/\rhoR$ are examined as labeled.
 }
\label{fig_EQprofile} 
\end{figure}

We adopt zero-beta MHD throughout, for which 
   the primitive quantities are the 
   mass density ($\rho$), velocity ($\vec{v}$), 
   and magnetic field ($\vec{B}$). 
Let the equilibrium quantities be denoted by a subscript $0$, 
   and consider only static equilibria ($\vec{v}_0 = 0$).
Let $(x, y, z)$ denote a Cartesian coordinate system,
   and let the uniform equilibrium magnetic field be $z$-aligned
   ($\vec{B}_0 = B_0 \uvec{z}$). 
We assume that the equilibrium density ($\rho_0$)
   depends only on $x$, following 
\begin{equation}
\label{eq_rhoEQ}
 \rho_0(x)
= 
 \left\{
   \begin{array}{ll}
   \rhoL, 											&            x<-\xext, \\ [0.2cm]
   \dfrac{1}{2}\left[(\rhoi+\rhoL)
	 -(\rhoi-\rhoL)\sin\dfrac{\pi(-x-d)}{l}\right],	& -\xext\le x\le -\xint, \\[0.2cm]
   \rhoi, 											& -\xint<x<\xint, \\[0.2cm]
   \dfrac{1}{2}\left[(\rhoi+\rhoR)
	-(\rhoi-\rhoR)\sin\dfrac{\pi(x-d)}{l}\right], 	& \xint\le x\le\xext, \\[0.2cm]
   \rhoR, 											& x>\xext,
\end{array}\right.
 \end{equation} 
	with 
\begin{equation}
  \xint = d-l/2, \quad 
  \xext = d+l/2.
\end{equation}	
Here $d$ represents some nominal slab half-width, and $l$
    the width of the two transition layers (TLs) that are geometrically
    symmetric about the nominal slab axis $x=0$. 
With subscript~${\rm i}$ we denote the equilibrium quantities 
    in the interior. 
Likewise, the subscript~${\rm L}$ (${\rm R}$) refers to the equilibrium values
    in the left (right) exterior.      
The \Alf\ speed is defined via $\va^2(x) = B_0^2/\mu_0 \rho_0(x)$ with $\mu_0$ being
   the magnetic permeability of free space. 
By $\vai$ and $\vaL$ ($\vaR$) we then mean the \Alf\ speeds 
   in the interior and left (right) exterior, respectively.  
Fixing $[\rhoi/\rhoL, l/d]$ at $[10, 0.5]$,      
   Figure~\ref{fig_EQprofile} plots $\rho_0$ against $x$ for several values 
   of $\rhoi/\rhoR$ as labeled. 
Two extreme configurations are relevant and displayed, 
   one corresponding to $\rhoR=\rhoL$ and the other to
   $\rhoR=\rhoi$. 
We consistently
   refer to the former (latter) as ``Fully Symmetric'' (``Fully Asymmetric'').

Oblique kink modes are in general resonantly absorbed in the \Alf\ continuum
   when the \Alf\ speed profiles are continuous 
   \citep[e.g., Section 8.14 in the textbook by][]{2019CUP_Roberts}, 
   a well-established fact that holds here despite the notion ``quasi-kink''.
We proceed with a resistive eigenmode approach
   (see the review by \citealt{2011SSRv..158..289G}, 
    hereafter \citetalias{2011SSRv..158..289G}, for conceptual clarifications).
Let the subscript $1$ denote small-amplitude perturbations, 
   which are governed by
\begin{eqnarray}
\label{eq_linMHD_momen}
 \rho_0 \dfrac{\partial \vec{v}_{1}}{\partial t}
&=&      \dfrac{(\nabla\times\vec{B}_1)\times \vec{B}_0}{\mu_0},
                    \\ 
\label{eq_linMHD_Farad}                     
 \dfrac{\partial \vec{B}_1}{\partial t}
&=& \nabla\times
         \left(\vec{v}_1\times \vec{B}_0-\dfrac{\eta}{\mu_0}\nabla\times\vec{B}_1
         \right). 
\end{eqnarray}
Here $\eta$ denotes the Ohmic resistivity, assumed
   to be constant for simplicity.
Any perturbation is Fourier-decomposed as
\begin{equation}
\label{eq_Fourier}
  g_1(x,y,z;t)=\Re\{\tilde{g}(x)\exp[-i(\Omega t-k_y y-k_z z)]\},
\end{equation}
   with $\Omega$ being the complex-valued eigenfrequency, 
   and $k_z$ ($k_y$) the real-valued axial (out-of-plane) wavenumber.
Let $\omega$ ($\gamma$) denote the real (imaginary) part of $\Omega$. 
Only damping eigensolutions are of interest ($\gamma<0$).   
The equations that further govern the Fourier amplitudes $\tilde{v}_x$, $\tilde{v}_y$,
   $\tilde{B}_x$, $\tilde{B}_y$, and $\tilde{B}_z$ 
   are identical to Equations~(6) to (10)
   in \citet[][hereafter \citetalias{2021SoPh..296...95Y}]{2021SoPh..296...95Y}.
As boundary conditions we require that 
   all Fourier amplitudes vanish far from the slab, given that only trapped modes
   are of interest.       
Following \citetalias{2021SoPh..296...95Y},     
   we formulate and solve the resulting EVP with
   the finite-element code PDE2D 
   \citep{1988Sewell_PDE2D}, which was first applied to 
   solar contexts by \citet{2006ApJ...642..533T} to our knowledge. 
We make sure that the details of the numerical setup,
   particularly where the boundaries are placed,    
   do not influence our numerical solutions.
The eigenfrequency $\Omega$ then formally writes
\begin{equation}
\label{eq_omgFormaltmp}
  \dfrac{\Omega d}{\vai}
= \mathcal{F}
  \left(\dfrac{\rhoi}{\rhoL},\dfrac{\rhoi}{\rhoR}, \dfrac{l}{d};
	    k_y d, k_z d; \Rm\right),
\end{equation}     
    where $\Rm=\mu_0 \vai d/\eta$ is some magnetic Reynolds number. 
Let asterisks denote complex conjugate.
The following symmetry properties then follow from the governing equations. 
If $\Omega$ is an eigenfrequency for a given pair $[k_y, k_z]$, 
   then so is $-\Omega^{*}$.
Furthermore, if $\Omega$ is an eigenfrequency for a given $[k_y, k_z]$,
   then it remains an eigenfrequency for $[-k_y,k_z]$, $[k_y,-k_z]$,
   and $[-k_y,-k_z]$.
One is therefore allowed to assume $\omega>0$ and consider only 
   the situation where $k_y \ge 0, k_z>0$.

\begin{figure}
\centering
 \includegraphics[width=.95\columnwidth]{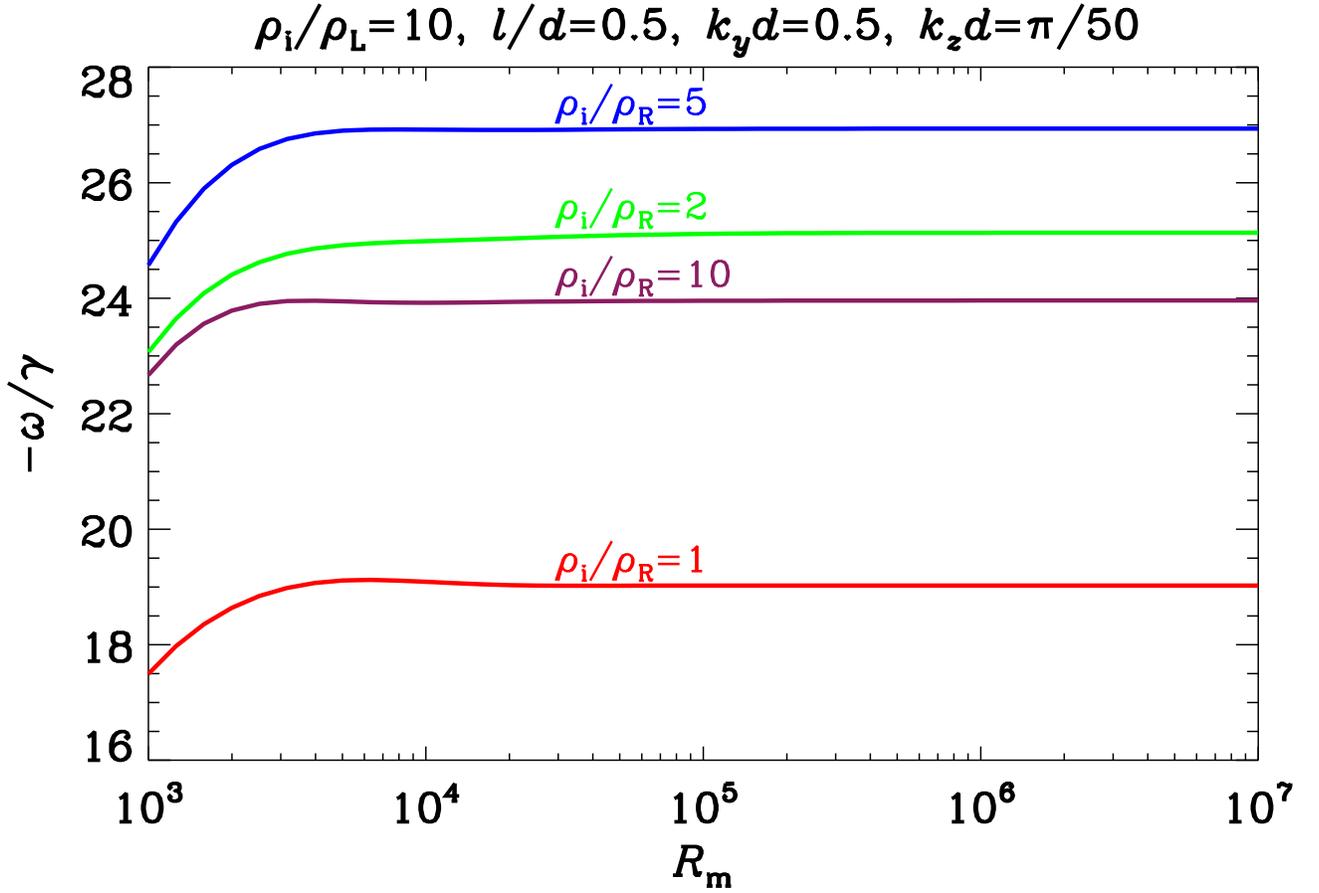}
 \caption{
Dependencies of the damping efficiency on the magnetic Reynolds number $\Rm$
   for oblique quasi-kink modes in asymmetric coronal slabs. 
The damping efficiency is measured by $\omega/|\gamma|$, with 
   $\omega$ being the oscillation frequency and $\gamma$ the damping rate. 
A fixed combination 
   $[\rhoi/\rhoL, l/d, k_y d, k_z d] = [10, 0.5, 0.5, \pi/50]$
   is adopted, whereas a number of values are examined
   for $\rhoi/\rhoR$ as labeled. 
 }
 \label{fig_eigen_Rm}
\end{figure}

Resonantly damped modes stand out in that their eigenfrequencies
    become $\Rm$-independent for sufficiently large $\Rm$.
This behavior was first shown by \citet{1991PhRvL..66.2871P} in fusion contexts,
    and later demonstrated for an extensive set of solar configurations
    \citep[e.g.,][]{2004ApJ...606.1223V,2006ApJ...642..533T,2016SoPh..291..877G,2018ApJ...868....5C,2021ApJ...908..230C}.
The same behavior is seen in Figure~\ref{fig_eigen_Rm}, where 
    the ratios of the oscillation frequency to the damping rate
    ($\omega/|\gamma|$) are plotted against the magnetic Reynolds number $\Rm$
    for a number of $\rhoi/\rhoR$ as labeled, with 
    the combination $[\rhoi/\rhoL, l/d, k_y d, k_z d]$ fixed at 
    $[10, 0.5, 0.5, \pi/50]$.
With Figure~\ref{fig_eigen_Rm} as an example, we quote a typical value 
    of $\sim 10^4-10^5$ for some critical $\Rm$ beyond which
    the eigenfrequencies remain constant. 
Only the $\Rm$-independent eigenfrequencies will be examined, meaning that
\begin{equation}
\label{eq_omgFormal}
  \dfrac{\Omega d}{\vai}
= \mathcal{F}
  \left(\dfrac{\rhoi}{\rhoL},\dfrac{\rhoi}{\rhoR}, \dfrac{l}{d};
	    k_y d, k_z d\right).
\end{equation}     
We further restrict ourselves, throughout this study,
    to those eigensolutions that are connected
    to the classic transverse fundamental kink mode
    arising in the situation where
    $\rhoL=\rhoR$, $l=0$ and $k_y=0$
    \citep[e.g., the textbook by][Figure~5.7]{2019CUP_Roberts}.
Evidently, the effects of density asymmetry can be brought out 
    by seeing $\rhoi/\rhoL$ as fixed 
	and examining those $\rhoi/\rhoR$ that are between
	       the ``Fully Symmetric'' ($\rhoi/\rhoR= \rhoi/\rhoL$)
	   and the ``Fully Asymmetric'' ($\rhoi/\rhoR=1$) limits. 
It then follows that the right \Alf\ continuum ($[k_z \vai, k_z \vaR]$)
   is always enclosed by the left one ($[k_z \vai, k_z \vaL]$).
Note that the right resonance is necessarily irrelevant (relevant)
    for a ``Fully Asymmetric'' (``Fully Symmetric'') configuration.
One therefore expects that the right resonance sets in only when
    $\rhoi/\rhoR$ exceeds some certain value when the rest of
    the parameters in the parentheses in Equation~\eqref{eq_omgFormal}
    are fixed.
To ease our description, by $x^{\rm A}_{\rm L}$ and $x^{\rm A}_{\rm R}$
    we consistently denote the locations of
    the left and right resonances even if the right resonance is absent. 
Regardless, we stress that the resonances are automatically handled by
    our resistive approach, and there is no need to consider the 
    relevance of $x^{\rm A}_{\rm R}$ beforehand. 
We further remark that an eigenfrequency $\Omega$ 
	is returned by the code together with the associated eigenfunctions, 
	the latter being
	dependent on $\Rm$ despite the $\Rm$-independence of the former.

\subsection{Energetics of Resonantly Damped Modes in Resistive MHD}
\label{sec_ProbFormu_sub_ener}

It turns out to be necessary to examine the small-amplitude perturbations
    from the energetics perspective.
We start by quoting a conservation law that follows from
    Equations~\eqref{eq_linMHD_momen} and \eqref{eq_linMHD_Farad}
    \citep[see e.g.,][for more general discussions]{1965RvPP....1..205B,1985GApFD..32..123L},     
\begin{eqnarray}
  \label{eq_linMHD_ener_cons}
  \dfrac{\partial \epsilon}{\partial t}
= -\nabla\cdot\vec{f}-\sres,
\end{eqnarray}
   in which
\begin{eqnarray}
  \epsilon
&=&  \dfrac{1}{2}\rho_0\vec{v}_1^2
    +\dfrac{\vec{B}_1^2}{2\mu_0},
			  \label{eq_linMHD_ener_den} \\
     \vec{f}
&=&
    \ptot \vec{v}_1
  - \dfrac{1}{\mu_0} (\vec{v}_1 \cdot \vec{B}_1) \vec{B}_0
  + \dfrac{\eta}{\mu_0} \vec{j}_1\times\vec{B}_1
 			  \label{eq_linMHD_f_final}	\\
   \sres 
&=&
  \eta \vec{j}_1^2.
			\label{eq_linMHD_sres_final} 		
\end{eqnarray}
Here $\ptot = \vec{B}_0\cdot\vec{B}_1/\mu_0$ 
    is the Eulerian perturbation of total pressure, 
    and $\vec{j}_1 = \nabla\times\vec{B}_1/\mu_0$
    is the perturbed electric current density.
Evidently, $\epsilon$ represents the instantaneous wave energy density,
    $\vec{f}$ the Poynting vector,
    and $\sres$ the Joule dissipation rate.

Suppose that a time-dependent system in question has settled to 
   a resonantly damped eigenmode. 
Let $\vec{k} = k_y\uvec{y}+k_z\uvec{z} = k\uvec{k}$
   denote a 2D wavevector, 
   with $k=|\vec{k}|$ and $\uvec{k}=\vec{k}/k$.
From $k$ we further define a wavelength $\lambda=2\pi/k$. 
One forms a right-handed coordinate system $(x, u_\perp, u_k)$ with $u_k$ being
   the coordinate in the $\vec{k}$-direction and $u_\perp$ the coordinate 
   in the third direction.
Now let $g_1$ and $h_1$ denote two arbitrary first-order quantities.    
It then follows from the Fourier ansatz (Equation~\eqref{eq_Fourier}) that
	$g_1$ or $h_1$ depends on $y$ and $z$ only via the term $k u_k = k_y y+k_z z$.  
Consequently, the average of the second-order quantity $g_1 h_1$
    over one wavelength $\lambda$ reads
\begin{equation}
\label{eq_def_ghAve}
    \left<g_1 h_1\right>(x, t) 
\coloneqq  \dfrac{1}{\lambda}\int_{0}^{\lambda}
    g_1(x, y, z, t) h_1(x, y, z, t) {\mathd}u_k
=   \overline{g_1 h_1}(x) \Exp{2\gamma t},
\end{equation}
   where 
\begin{equation}
\label{eq_def_overbar}
  \overline{g_1 h_1}(x) 
=
  \dfrac{1}{2} {\Re}[\tilde{g}^*(x)\tilde{h}(x)]
= \dfrac{1}{2} {\Re}[\tilde{g}(x)  \tilde{h}^*(x)].  
\end{equation} 
We proceed with the well known fact that 
    dissipative effects are important only in
    some dissipation layers (DLs) that embrace the resonances 
    (see \citetalias{2011SSRv..158..289G} and references therein).
Let $[x^{-}_{\rm L}, x^{+}_{\rm L}]$ ($[x^{-}_{\rm R}, x^{+}_{\rm R}]$) 
	denote the left (right) DL.
Consider a fixed volume $V$ that spans a length
    of $\lambda$ in the $\vec{k}$-direction
    and is of unit length in the $u_\perp$-direction. 
Furthermore, let its $x$-extent be the entire $x$-axis with the exception
    of the two DLs.
Taking $\eta=0$ and integrating Equation~\eqref{eq_linMHD_ener_cons} 
    over $V$, one finds by repeatedly using Equation~\eqref{eq_def_ghAve} 
    that
\begin{equation}
\label{eq_enerCons_1D}
-2\gamma \hat{E} = \hat{F}_{\rm L} + \hat{F}_{\rm R},
\end{equation}    
    where
\begin{equation}
\label{eq_def_EF}
\begin{split}
& \hat{E} = \left(\int_{      -\infty}^{x^{-}_{\rm L}}
         +\int_{x^{+}_{\rm L}}^{x^{-}_{\rm R}}
         +\int_{x^{+}_{\rm R}}^{\infty}\right)\bar{\epsilon}(x){\mathd}x, \\
& \hat{F}_{\rm L} =\bar{f}_x(x^{-}_{\rm L}) -\bar{f}_x(x^{+}_{\rm L}),    \\
& \hat{F}_{\rm R} =\bar{f}_x(x^{-}_{\rm R}) -\bar{f}_x(x^{+}_{\rm R}). 
\end{split}
\end{equation}   
Moreover, $f_x = \ptot v_{1x}$ is the $x$-component of the Poynting vector.
Technical details aside, Equation~\eqref{eq_enerCons_1D}
    reflects the simple fact that the wave energy in the ideal portions
    of the system is lost only via the net energy flux into
    the DLs where the \Alf\ resonances take place.        
It therefore follows that the contributions of individual resonances
    to the gross damping rate ($\gamma$) are measured 
    by $\hat{F}_{\rm L}/2\hat{E}$ and $\hat{F}_{\rm R}/2\hat{E}$.
Evidently, $\hat{F}_{\rm R} = 0$ when the right resonance does not occur.      

\section{Oblique Quasi-kink Modes in the Thin-Boundary Limit}
\label{sec_TB}
This section makes some analytical progress in
    the thin-boundary (TB) limit ($l/d\ll 1$) by
    capitalizing on the formulations for generic 1D equilibria
	\citep[for first derivations, see e.g.,][]
	{1991SoPh..133..227S,1992SoPh..138..233G,1995JPlPh..54..129R,1996ApJ...471..501T}.
Our purposes are twofold.
Firstly, we will derive the relevant dispersion relation (DR) such that 
    its solutions can be employed to validate our resistive computations.
We deem this validation necessary given that oblique quasi-kink modes have
    not been examined when the configuration does not take
    the ``Fully Symmetric'' or ``Fully Asymmetric'' limit.       
Secondly, we will collect some analytical expressions that approximately
    solve the DR for the two limiting configurations.
This proves necessary not only for validation purposes but for understanding our
    numerical results on how the density asymmetry influences the
    phase and group diagrams.      

\subsection{General Formulations}
\label{sec_TB_sub_genFormu}	
We start by noting that the ideal version ($\eta=0$) of the governing equations
    can be combined to yield a well known equation for $\tilde{v}_x$
    (e.g., \citealt{2007SoPh..246..213A}; \citetalias{2021SoPh..296...95Y};
    and references therein)
\begin{equation}
\label{eq_2nd_order_vx}  
  \left[
  	   \dfrac{k_z^2 - \Omega^2/\va^2}{k_y^2 + k_z^2 - \Omega^2/\va^2}
    	   \tilde{v}_x'
  \right]'
- \left(k_z^2 - \Omega^2/\va^2\right)\tilde{v}_x
= 0, 
\end{equation}
    where the shorthand notation $' = \mathd/\mathd x$ is employed. 
The \Alf\ resonance takes place wherever $\Omega = k_z \va$, 
    and turns out to always (not necessarily) arise in the left (right) TL.  
Regardless, we consistently label the resonance(s) with 
    the superscript ${\rm A}$, which is supplemented with the subscripts 
    ${\rm L}$ or ${\rm R}$ when the left and right resonances
    need to be discriminated. 
We proceed by defining 
\begin{equation}
\label{eq_def_KappaM}
\kappa_{j}^2 = k_z^2 - \dfrac{\Omega^2}{v_{{\rm A}j}^2}, \quad  
m_{j}^2      = k_y^2 + \kappa_{j}^2  
  	         = k_y^2 + k_z^2 - \dfrac{\Omega^2}{v_{{\rm A}j}^2}, 
\end{equation}
    where $j={\rm i, L, R}$ and we take 
    $-\pi/2 < \arg\kappa_{j}, \arg m_{j} \le \pi/2$ without loss of generality.  
The solution to Equation~\eqref{eq_2nd_order_vx} in the uniform regions 
    then writes
\begin{equation}
\label{eq_SolTB_vx}
  \tilde{v}_x(x)
= \left\{
  \begin{array}{ll}
   A_1\exp(\mL x), 					&  x<-\xext,      \\ [0.1cm]
   C_1\cosh(\mi x)+C_2\sinh(\mi x), & -\xint<x<\xint, \\ [0.1cm]
   A_2\exp(-\mR x), 				&  x>\xext,
 \end{array}
    \right.
 \end{equation}
   with $A_{1, 2}$ and $C_{1,2}$ being constants.
Likewise, the Fourier amplitude of the Eulerian perturbation of total pressure
   is given by
\begin{equation}
\label{eq_SolTB_pTot}
  \tilde{p}_{\rm tot}
=-\dfrac{i}{\Omega} \dfrac{B_0^2}{\mu_0}
  \times
  \left\{
    \begin{array}{ll}
     \dfrac{\kappaL^2}{\mL}A_1\exp(\mL x), 			 &  x<-\xext,      \\ [0.4cm]
     \dfrac{\kappai^2}{\mi}
       \left[C_1\sinh(\mi x)+C_2\cosh(\mi x)\right], & -\xint<x<\xint, \\ [0.4cm]
    -\dfrac{\kappaR^2}{\mR}A_2\exp(-\mR x), 		 &  x>\xext.
    \end{array}
  \right.
\end{equation} 

The relevant DR in the TB limit is derived as follows. 
By construction, a DL where dissipative effects 
    are important is thin, bracketing a resonance and bracketed by
    a TL (see e.g., \citetalias{2011SSRv..158..289G} for technical details). 
Let the variation of some quantity $q$
    across a DL be denoted by $\llbracket q \rrbracket$, which is further taken
    by the TB treatment to be the variation of $q$ across the pertinent TL. 
The end result is that \citep[e.g.,][]{1996ApJ...471..501T,2000ApJ...531..561A}
\begin{equation}
\label{eq_TB_connec}
\begin{split}
& \llbracket \tilde{p}_{\rm tot} \rrbracket = 0,  \\[0.2cm]
& \llbracket \tilde{\xi}_x       \rrbracket 
 = -\dfrac{i\pi k_y^2 {\rm sgn}(\omega)}{\rho^{\rm A}|\Delta^{\rm A}|}
   \tilde{p}^{\rm A}_{\rm tot},
\end{split}
\end{equation}
   where $\tilde{\xi}_x$ represents the Fourier amplitude of the transverse
   Lagrangian displacement defined via $\tilde{v}_x = -i\Omega \tilde{\xi}_x$.
By the superscript ${\rm A}$ we mean that 
    the relevant quantity is evaluated at a resonance
    ($x^{\rm A}_{\rm L}$ or $x^{\rm A}_{\rm R}$).
In particular, $\Delta^{\rm A}$ is defined by
    \citep[see][where it was first introduced]{1991SoPh..133..227S}
\begin{equation}
\label{eq_TB_defDelta}
  \Delta^{\rm A} 
= \left.
  \dfrac{\mathd(\omega^2-k_z^2 \va^2)}{\mathd x}\right|_{x=x^{\rm A}}.
\end{equation}   
A DR results when one connects the solutions
   in the uniform regions (Equations~\eqref{eq_SolTB_vx} and~\eqref{eq_SolTB_pTot})
   by the connection formulas \eqref{eq_TB_connec}, reading 
\begin{equation}
\label{eq_TB_DR}
\begin{split}
 & \left(
      \dfrac{\kappaL^2}{\mL}\dfrac{\kappaR^2}{m_{\rm R}}
     +\dfrac{\kappai^4}{\mi^2}
   \right)
  +\left(\dfrac{\kappaL^2}{\mL}+\dfrac{\kappaR^2}{\mR}\right)
   \dfrac{\kappai^2}{\mi}
   \coth(2\mi d)
  +\dfrac{\pi^2 k_y^4}{k_z^4}
   \dfrac{\kappaL^2}{\mL}
   \dfrac{\kappaR^2}{\mR}
   \dfrac{\kappai^4}{\mi^2}
   \dfrac{\rho^{\rm A}_{\rm L}}{(\rho')^{\rm A}_{\rm L}}
   \dfrac{\rho^{\rm A}_{\rm R}}{(\rho')^{\rm A}_{\rm R}} \\[0.5cm]
 & +
  \dfrac{i\pi k_y^2}{k_z^2}
  \dfrac{\kappaL^2}{\mL}
  \dfrac{\kappaR^2}{\mR}
  \dfrac{\kappai^2}{\mi}
  \left[\dfrac{\rho^{\rm A}_{\rm L}}{(\rho')^{\rm A}_{\rm L}}
       -\dfrac{\rho^{\rm A}_{\rm R}}{(\rho')^{\rm A}_{\rm R}}
       \right] 
  \coth(2\mi d)
  +
  \dfrac{i\pi k_y^2}{k_z^2}
  \dfrac{\kappai^4}{\mi^2}
  \left[
       \dfrac{\rho^{\rm A}_{\rm L}}{(\rho')^{\rm A}_{\rm L}}
  	   \dfrac{\kappaL^2}{\mL}
  	 - \dfrac{\rho^{\rm A}_{\rm R}}{(\rho')^{\rm A}_{\rm R}}
  	   \dfrac{\kappaR^2}{\mR}   
  \right]
=0, 
\end{split}
\end{equation}   
    where $\rho$ actually means the equilibrium density $\rho_0$.
Note that the identity $\rho_0(x)\va^2(x) = {\rm const}$ in zero-beta MHD 
    is employed to slightly simplify Equation~\eqref{eq_TB_DR}.
Note further that the right resonance is assumed to occur,
    as represented by the symbols $\rho^{\rm A}_{\rm R}$ and $(\rho')^{\rm A}_{\rm R}$.
We have additionally seen $\omega$ as positive, and employed
    the fact that $(\rho')^{\rm A}_{\rm L} >0$ and $(\rho')^{\rm A}_{\rm R} <0$ 
    given our density profile (Equation~\eqref{eq_rhoEQ}).

Some remarks are necessary here. 
Firstly, so far the examinations on quasi-kink modes in an asymmetric slab system 
    pertain exclusively to the case where $k_y = 0$ and $l=0$
    \citep[e.g.,][]{2017SoPh..292...35A, 2018ApJ...853..136K,2020ApJ...894..123Z}.
Take the study by \citet{2018ApJ...853..136K}.
The zero-beta version of Equation~(16) therein writes  
\begin{equation}
\label{eq_TB_DR_ky0}
     2(\kappai^2+\kappaL\kappaR) 
   + \kappai(\kappaL+\kappaR)
       [\tanh(\kappai d)+\coth(\kappai d)] = 0
\end{equation}
    with our notations. 
One readily verifies that Equation~\eqref{eq_TB_DR_ky0} is recovered
    by our Equation~\eqref{eq_TB_DR} when $k_y=0$. 
Generally speaking, neither Equation~\eqref{eq_TB_DR} nor Equation~\eqref{eq_TB_DR_ky0} 
    can be factorized given the coupling between
    kink-like and sausage-like motions. 
Secondly, our discussions on Equation~\eqref{eq_omgFormal} 
    indicate that the right resonance sets in only when
    $\rhoi/\rhoR$ exceeds some critical value $(\rhoi/\rhoR)_{\rm crit}$.
While assuming the relevance of the right resonance,
    Equation~\eqref{eq_TB_DR} can actually account for the situation 
    where $\rhoi/\rhoR < (\rhoi/\rhoR)_{\rm crit}$ by simply 
    letting $\rho^{\rm A}_{\rm R}/(\rho')^{\rm A}_{\rm R} = 0$.  
One complication, however, is that in general we do not know when to switch off 
    the $\rho^{\rm A}_{\rm R}/(\rho')^{\rm A}_{\rm R}$ terms beforehand.
Given our purposes,     
    we choose to solve Equation~\eqref{eq_TB_DR} for only 
    the two limiting cases ($\rhoi/\rhoR=\rhoi/\rhoL$ and $\rhoi/\rhoR=1$)
    plus one value of $\rhoi/\rhoR$ that lies in between. 
We discard the right resonance only when $\rhoi/\rhoR=1$.  
The range of $l/d$, on the other hand, is taken to be rather broad.        
Regardless, Equation~\eqref{eq_TB_DR} is always solved in an iterative manner.
With a guess for $\Omega$, we determine the resonance location(s),
    evaluate the terms with a superscript ${\rm A}$, and then update $\Omega$
    by a standard root-finder.     
This process is repeated until convergence. 
Note that the intermediate value of $\rhoi/\rhoR$ is chosen such that
    the right resonance arises for the entire range of $l/d$ to be examined.
Note further that the full form of Equation~\eqref{eq_TB_DR} is solved 
    for the two limiting cases, 
    despite its simplifications in what follows.

	
\subsection{``Fully Symmetric'' and ``Fully Asymmetric'' Configurations}
\label{sec_TB_sub_SymAsym}		
Consider first a ``Fully Symmetric'' configuration ($\rhoR = \rhoL$).
It immediately follows from Equation~\eqref{eq_def_KappaM} 
   that $\kappaL = \kappaR$ and $\mL = \mR$. 
The right resonance is guaranteed,
   satisfying the relations $x^{\rm A}_{\rm R} = -x^{\rm A}_{\rm L}$, 
   $\rho^{\rm A}_{\rm R} = \rho^{\rm A}_{\rm L}$,
   and $(\rho')^{\rm A}_{\rm R} = -(\rho')^{\rm A}_{\rm L}$.
Defining
\begin{equation}
\label{eq_TB_defAlpha}
   \alpha 
=       \dfrac{\kappai^2}{\mi}\dfrac{\mL}{\kappaL^2}
  -i\pi \dfrac{k_y^2}{k_z^2} 
  		\dfrac{\kappai^2}{\mi}
  		\dfrac{\rho^{\rm A}_{\rm L}}{(\rho')^{\rm A}_{\rm L}},
\end{equation}
    one finds that the left hand side (LHS) of Equation~\eqref{eq_TB_DR} can be
    factorized,
     the eigenfrequency $\Omega$ satisfying
     either $\coth(\mi d)+\alpha = 0$ or $\tanh(\mi d)+\alpha = 0$.
The former is the DR for resonantly damped oblique kink modes
	\footnote{The latter governs oblique sausage modes, which are beyond
	our scope here. We remark only that this relation becomes
	$\tanh(\mi d)=-(\kappai^2/\mi)(\mL/\kappaL^2)$ when $l=0$, 
	thereby recovering, say, 
	Equation~(13) in \citealt{2007SoPh..246..213A}.}, 
    and more explicitly writes 
\begin{equation}
\label{eq_TB_FullSymDR}
   \coth(\mi d) 
= -\dfrac{\kappai^2}{\mi}\dfrac{\mL}{\kappaL^2}
  +i\pi \dfrac{k_y^2}{k_z^2} 
  		\dfrac{\kappai^2}{\mi}
  		\dfrac{\rho^{\rm A}_{\rm L}}{(\rho')^{\rm A}_{\rm L}}.
\end{equation}    
The first derivation of Equation~\eqref{eq_TB_FullSymDR} with 
    the resistive eigenmode approach was due to \citet{1992SoPh..138..233G}. 
Relevant here is the situation 
    where $\mi^2 \approx \mL^2\approx k_y^2$ and $|\gamma|\ll\omega$.
Specializing to our density profile (Equation~\eqref{eq_rhoEQ}),    
    the approximate solution to Equation~\eqref{eq_TB_FullSymDR} 
    can be summarized as
\begin{eqnarray}
&&        \omega^2 
  \approx k_z^2 \vai^2
  		 \dfrac{1+\Theta}{\rhoL/\rhoi+\Theta},  
   		 \label{eq_TB_FullSym_omgR}\\
&&  \dfrac{\gamma}{\omega}
\approx 
   -\left(k_y l\right) 
    \left(\dfrac{1+\Theta}{2\sqrt{\Theta}}\right)
    \left[\dfrac{(1-\rhoL/\rhoi)\Theta^2}{(\rhoL/\rhoi+\Theta)(1+\Theta)^2}\right],
    	\label{eq_TB_FullSym_omgI}
\end{eqnarray}    
	provided 
\begin{equation}
\label{eq_TB_FullSym_valid}
	k_y^2 (\rhoL/\rhoi+\Theta) \gg k_z^2 (1-\rhoL/\rhoi), \quad
	{\rm and} \quad
	|\gamma| \ll \omega.  
\end{equation}
Here $\Theta = \tanh(k_y d)$.	
Equations~\eqref{eq_TB_FullSym_omgR} and \eqref{eq_TB_FullSym_omgI} 
	were given in \citetalias{2021SoPh..296...95Y}, 
	and a slightly different version was first derived for some different
	density profile by \citet{1998JPSJ...67.2322T} in fusion contexts. 
The first inequality in Equation~\eqref{eq_TB_FullSym_valid} follows from
    the requirement $\mi^2 \approx \mL^2 \approx k_y^2$, and is derived here
    to make clearer the range of validity. 
The approximate solution
    simplifies considerably if one further assumes $k_y d \gg 1$
    (namely $\Theta\approx 1$), a case that has been much-studied
    \citep[e.g.,][to name only a few]{1978ApJ...226..650I,1992SoPh..138..233G,1995JPlPh..54..129R}.
In particular, Equation~\eqref{eq_TB_FullSym_omgR} becomes
\begin{equation}
\label{eq_TB_def_ck}
\omega^2 \approx k_z^2 \ckL^2 = k_z^2 \dfrac{2\vai^2}{\rhoL/\rhoi+1},
\end{equation}   
    with $\ckL$ being the classic kink speed
    despite the cumbersome subscript ${\rm L}$. 
    
Now move on to a ``Fully Asymmetric'' configuration ($\rhoR = \rhoi$),
   for which the right TL and hence the right resonance are absent.
The relevant DR is actually also a special case of Equation~\eqref{eq_TB_DR},
   provided that one takes $\rho^{\rm A}_{\rm R}/(\rho')^{\rm A}_{\rm R} = 0$
   and does not discriminate $\kappaR^2/\mR$ from $\kappai^2/\mi$.
Eliminating a common factor $1+\coth(2\mi d)$, one finds that the DR writes 
\begin{equation}
\label{eq_TB_FullAsym}
  \dfrac{\kappaL^2}{\mL}+\dfrac{\kappai^2}{\mi}
+ \dfrac{i\pi k_y^2}{k_z^2}
  \dfrac{\kappaL^2}{\mL}
  \dfrac{\kappai^2}{\mi}
  \dfrac{\rho^{\rm A}_{\rm L}}{(\rho')^{\rm A}_{\rm L}}
= 0,
\end{equation}
  where the cumbersome subscript ${\rm i}$
  is retained instead of  ${\rm R}$
  to maintain formal consistency. 
Note that this configuration, pertinent to sheet pinch in fusion contexts,
    was the one that led to the concept of spatial resonance
    \citep{1973ZPhy..261..203T,1973ZPhy..261..217G,1974PhRvL..32..454H,1974PhFl...17.1399C}.
There have been numerous solar applications of this configuration as well
	\citep[e.g.,][to name only a few early studies]{1979ApJ...233..756W,1986ApJ...301..430L,1988JGR....93.5423H},
	and the approximate expressions of interest
	($\mi^2\approx\mL^2\approx k_y^2$, $|\gamma|\ll\omega$) can be collected as
\begin{eqnarray}
&& \omega^2 
   \approx k_z^2 \ckL^2, 
		\label{eq_TB_FullAsym_omgR}\\
&& \dfrac{\gamma}{\omega}
   \approx 
   -\dfrac{k_y l}{4}\dfrac{1-\rhoL/\rhoi}{1+\rhoL/\rhoi},
   		\label{eq_TB_FullAsym_omgI}
\end{eqnarray}
   which holds for our density profile (Equation~\eqref{eq_rhoEQ})
   provided that
\begin{equation}
\label{eq_TB_FullAsym_valid}
k_y^2 (\rhoL/\rhoi+1) \gg k_z^2 (1-\rhoL/\rhoi), \quad
	{\rm and} \quad
	|\gamma| \ll \omega.  
\end{equation}    
Equations~\eqref{eq_TB_FullAsym_omgR} and \eqref{eq_TB_FullAsym_omgI},
    together with their explicit range of validity
    (Equation~\eqref{eq_TB_FullAsym_valid}), 
    are seen to agree with the $\Theta\approx 1$ limit of
    the ``Fully Symmetric'' results.

\section{Results I: Rates of Resonant Absorption}
\label{sec_ResIRA}
This section examines the resonant damping of oblique quasi-kink modes, 
    assuming that $k_y$ and $k_z$ can be observationally identified. 
Evidently, the parentheses in Equation~\eqref{eq_omgFormal} contain
    way too many parameters to exhaust.
We choose a fixed $\rhoi/\rhoL=10$,  
    a density contrast that is reasonable for say,
    AR loops \citep[e.g.,][]{2004ApJ...600..458A},
    polar plumes \citep[e.g.,][]{2011A&ARv..19...35W},
    and streamer stalks \citep[e.g.,][]{2011ApJ...728..147C}.   
The dimensionless axial wavenumber $k_z d$ will be fixed at $\pi/50$,
    which is reasonable for axial fundamentals or their first several harmonics in coronal structures
    typically imaged in the EUV
    \citep[e.g.,][Figure~1]{2007ApJ...662L.119S}.          
We additionally restrict ourselves to the situation
    where $k_y > k_z$ (or even $k_y \gg k_z$), largely organizing our results
    around the role of $\rhoi/\rhoR$.

\subsection{Validation of Resistive Computations Against Thin-Boundary Expectations}
\label{sec_ResIRA_subValid}

\begin{figure}
\centering
 \includegraphics[width=.95\columnwidth]{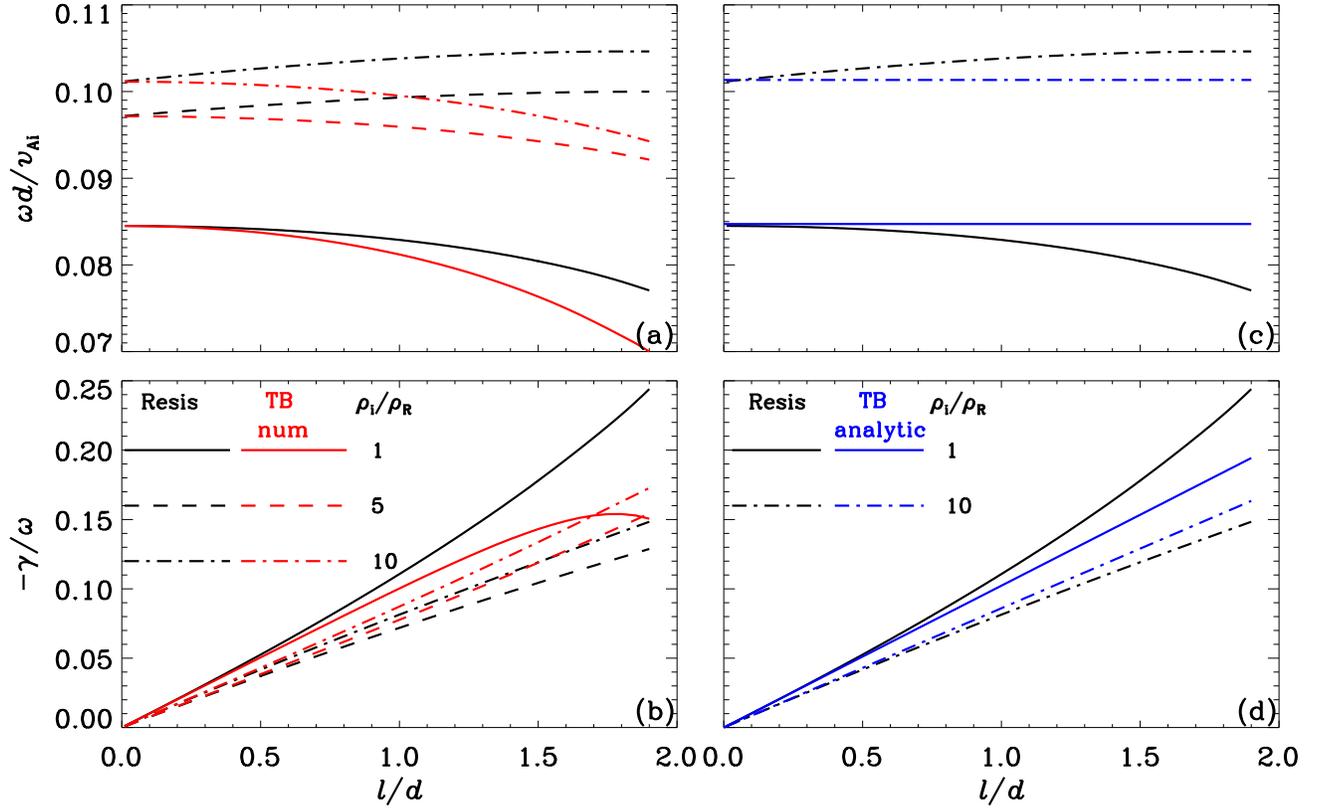}
 \caption{
 Dispersion properties of resonantly damped oblique
     quasi-kink modes in an asymmetric slab configuration. 
 Plotted here are the oscillation frequency ($\omega$, the upper row)
      and the ratio of the damping rate to the oscillation frequency
      ($-\gamma/\omega$, lower) against the dimensionless TL width ($l/d$). 
 A fixed combination $[\rhoi/\rhoL, k_y d, k_z d] = [10, 0.5, \pi/50]$
    is adopted, whereas a number of $\rhoi/\rhoR$ are examined
    as discriminated by the line styles. 
 The black curves represent our resistive computations (labeled ``Resis'').
 Also presented are two groups of solutions to the
   dispersion relation (Equation~\eqref{eq_TB_DR}) 
   in the thin-boundary limit, 
   one found numerically
     (labeled ``TB num'', the red curves in the left column)
   and the other found analytically in an approximate manner
     (``TB analytic'', the blue curves in the right column).
 See text for more details.      
 }
 \label{fig_cpTB}
\end{figure}

Figure~\ref{fig_cpTB} starts our examination by comparing 
    the resistive results (labeled ``Resis'', the black curves)
    with the relevant
    TB expectations (red and blue) for 
    a fixed combination $[\rhoi/\rhoL, k_y d, k_z d] = [10, 0.5, \pi/50]$.
Plotted are the oscillation frequency ($\omega$, the upper row)
    and the ratio of the damping rate to the oscillation frequency
    ($-\gamma/\omega$, lower) as functions of the dimensionless TL width ($l/d$).
A number of values are examined for $\rhoi/\rhoR$ as discriminated by
    the line styles. 
Two groups of TB results are presented, one being
    the iterative solutions to Equaiton~\eqref{eq_TB_DR}
     (labeled ``TB num'', the left column),
    and the other being those evaluated with the approximate analytical expressions
    (``TB analytic'', right).
Note that only the ``Fully Asymmetric'' ($\rhoi/\rhoR=1$)
    and ``Fully Symmetric'' ($\rhoi/\rhoR=\rhoi/\rhoL$)
    cases are examined in the right column, given the limited availability
    of the analytical expressions 
    (see Equations~\eqref{eq_TB_FullSym_omgR}
         and \eqref{eq_TB_FullSym_omgI} 
         as well as \eqref{eq_TB_FullAsym_omgR}
         and \eqref{eq_TB_FullAsym_omgI}). 
Consider the left column first.
One immediately sees that the resistive results agree remarkably well with the 
    ``TB num'' ones for, say, $l/d \lesssim 1$, meaning that the two independent
    approaches are both correctly implemented. 
Furthermore, there actually exists a rather close agreement between
    the two sets of solutions even for $l/d \sim 2$, the only exception being
    in the $-\gamma/\omega$ profile for $\rhoi/\rhoR=1$.
Now move on to the right column, where one sees that
    the approximate expressions perform even better
    than the iterative solutions in reproducing the resistive results, 
    despite that the iterative approach is more self-consistent in principle. 
Overall, Figure~\ref{fig_cpTB} offers yet another piece of evidence that
    the TB expectations for coronal equilibria may hold well beyond
    their nominal range of applicability 
    \citep[e.g.,][]{2004ApJ...606.1223V,2013ApJ...777..158S,2021ApJ...908..230C}.
We conclude further that faith can be placed in our resistive approach,
    which will be consistently adopted hereafter.
As for the TB approximation, we choose to invoke only
    the approximate expressions for the two limiting values of $\rhoi/\rhoR$
    when necessary.
The reasons for us to do this are largely twofold, one being the complication
    for assessing the relevance of the right resonance a priori,
    and the other being the general difficulty to further proceed analytically
    when $\rhoi/\rhoR$ is between the extreme values.

\subsection{Effects of Density Asymmetry} 
\label{sec_ResIRA_subDenAsym}
\begin{figure}
\centering
 \includegraphics[width=.85\columnwidth]{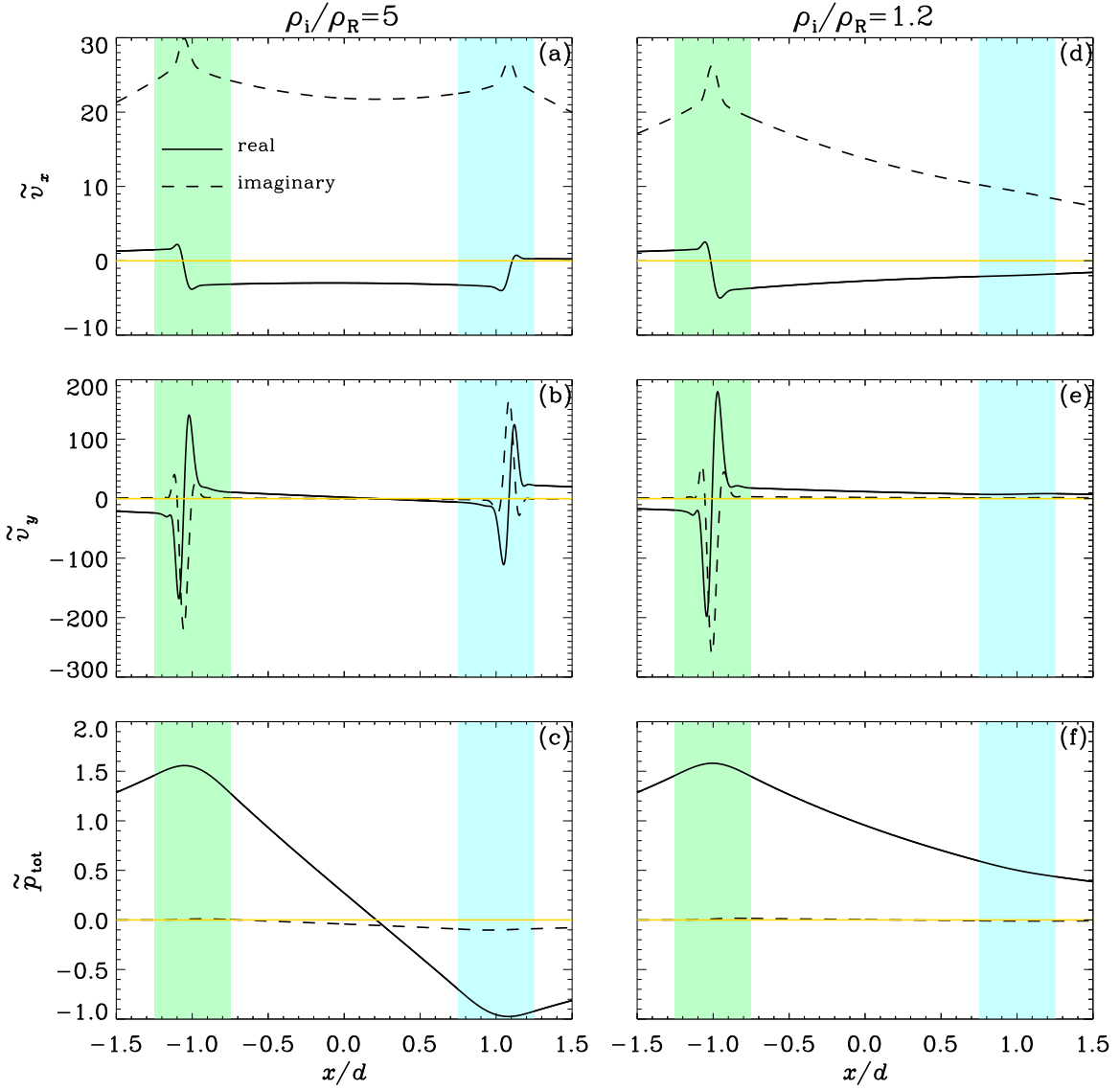}
 \caption{
 Eigenfunctions of resonantly damped oblique quasi-kink modes 
    in an asymmetric slab configuration.
 A fixed combination $[\rhoi/\rhoL, l/d, k_y d, k_z d]=[10, 0.5, 0.5, \pi/50]$ 
    is adopted, whereas two values are examined for $\rhoi/\rhoR$,
    one being $5$ (the left column)
        and the other being $1.2$ (right).  
 The portions shaded green and blue correspond to the
    left and right transition layers, respectively.
 The magnetic Reynolds number $\Rm$ is taken to be $10^5$.         
 Plotted from top to bottom are the Fourier amplitudes of
    the transverse speed ($\tilde{v}_x$),
    the out-of-plane speed ($\tilde{v}_y$),
    and the total pressure perturbation ($\tilde{p}_{\rm tot}$).
 The eigenfunctions are scaled such that 
    $\tilde{p}_{\rm tot}=1$ at $x=-2d$, 
    their real (imaginary) parts represented by the solid (dashed) curves.
 }
 \label{fig_eigFun}
\end{figure}

Whether the right resonance occurs is most readily revealed
    by the spatial profiles of the resistive eigenfunctions, for which
    purpose Figure~\ref{fig_eigFun} plots the Fourier amplitudes of
    the transverse speed ($\tilde{v}_x$, the top row),
    the out-of-plane speed ($\tilde{v}_y$, middle),
    and the Eulerian perturbation of total pressure ($\tilde{p}_{\rm tot}$, bottom).
Two values for $\rhoi/\rhoR$ are examined, one being $5$ (the left column)
    and the other being $1.2$ (right), whereas  
    the combination $[\rhoi/\rhoL, l/d, k_y d, k_z d]$ is fixed at
    $[10, 0.5, 0.5, \pi/50]$.
We additionally take the magnetic Reynolds number $\Rm$
    to be $10^5$ for both columns.     
The left and right TLs correspond to the portions shaded green and blue,
    respectively.
Furthermore, the eigenfunctions are scaled such that 
    $\tilde{p}_{\rm tot}$ attains unity at $x=-2d$, 
    their real (imaginary) parts represented by the solid (dashed) curves.
Now that ideal MHD essentially applies at $x=-2d$, 
    our way for rescaling the eigenfunctions means that any \Alf\ 
    resonance is characterized by the following features 
    (see \citetalias{2011SSRv..158..289G} and references therein).
The strongest dynamics occurs for $\tilde{v}_y$ given the so-called
    $1/s$ singularity, and the dynamics of $\Im\tilde{v}_x$
    is the second strongest as a result of some $\ln s$ singularity. 
The total pressure perturbation $\tilde{p}_{\rm tot}$
    possesses the least strong variation, which in fact cannot be discerned
    for the examined parameters.     
One therefore sees that the right resonance is relevant (irrelevant)
    when $\rhoi/\rhoR=5$ ($\rhoi/\rhoR=1.2$).
With the left resonance for $\rhoi/\rhoR=5$ as an example,
    one further sees that $\Re\tilde{v}_x$ jumps at any resonance location
    (see e.g., Figure~\ref{fig_eigFun}a),
    and $\tilde{p}_{\rm tot}$ is dominated by its real part there
    (e.g., Figure~\ref{fig_eigFun}c). 
Given Equation~\eqref{eq_def_EF}, the sign of
    $\Re\tilde{p}_{\rm tot}$ and that of the jump in $\Re\tilde{v}_x$  
    then dictate that $\hat{F}$, the net energy flux into a resonance,
    is always positive.
On top of that, an inspection of
    the magnitudes of $\Re\tilde{p}_{\rm tot}$
    and the jump in $\Re\tilde{v}_x$ in the left column
    indicates that $\hat{F}_{\rm L}>\hat{F}_{\rm R}$, meaning that
    the left resonance plays a more important role in damping 
    the oblique quasi-kink mode at hand.    

\begin{figure}
\centering
 \includegraphics[width=.7\columnwidth]{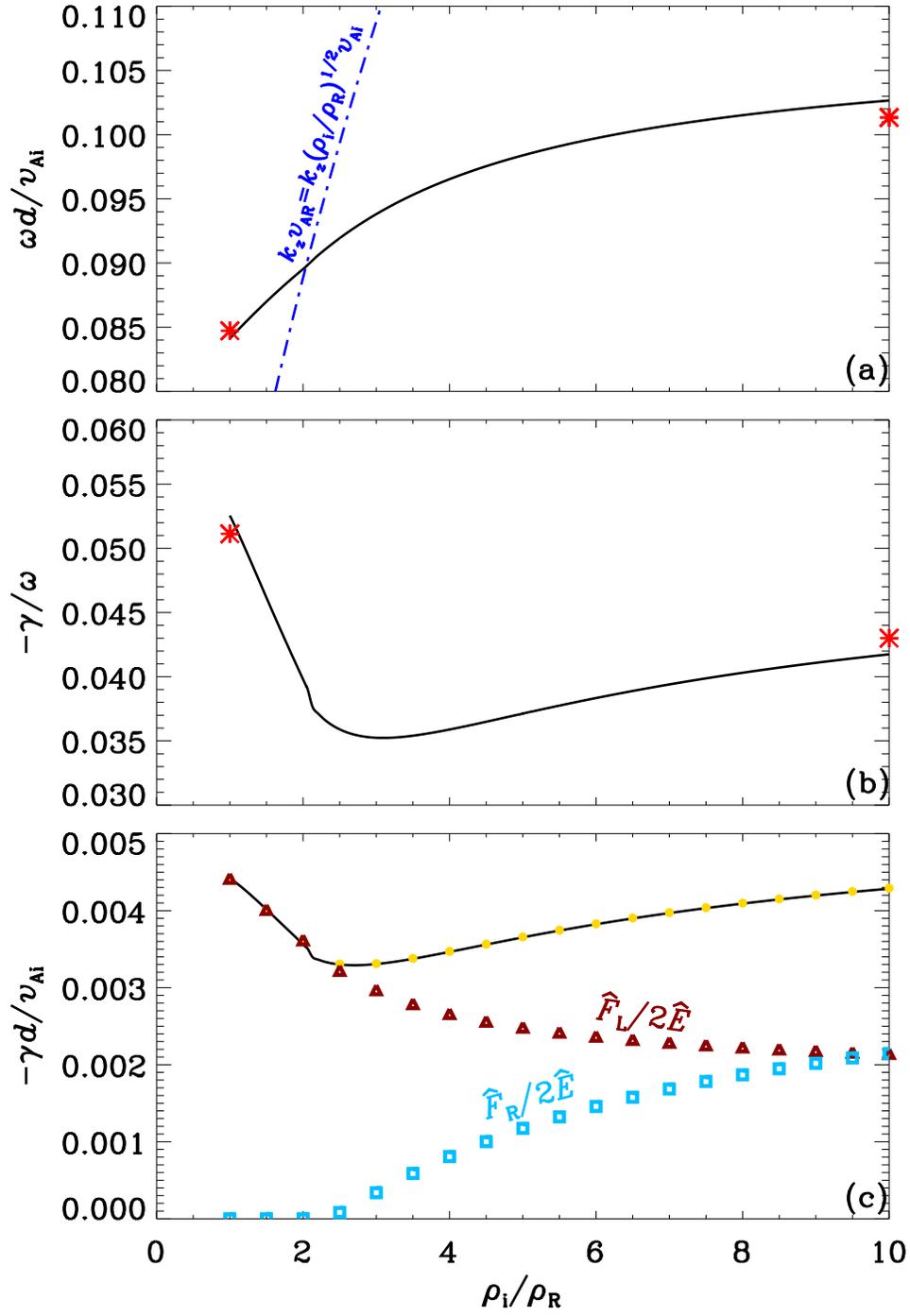}
 \caption{
Effects of density asymmetry on resonantly damped oblique quasi-kink modes
    in an asymmetric slab configuration. 
Plotted are the dependencies on $\rhoi/\rhoR$ of 
    (a) the oscillation frequency $\omega$,
    (b) the ratio of the damping rate to oscillation frequency $-\gamma/\omega$,        
    and (c) the damping rate $-\gamma$. 
The combination $[\rhoi/\rhoL, l/d, k_y d, k_z d]$
    is fixed at $[10, 0.5, 0.5, \pi/50]$.    
The red asterisks in (a) and (b) 
    are obtained with the approximate expressions
    for the ``Fully Asymmetric'' and ``Fully Symmetric'' configurations
    in the thin-boundary limit. 
Furthermore, the blue dash-dotted curve in (a) represents the upper end
    of the right \Alf\ continuum ($k_z \vaR$). 
The open triangles (boxes) in (c) correspond to the contribution
    to the gross damping rate from the left (right) resonance,
    as evaluated with the relevant eigenfunctions. 
The sum of the individual contributions 
    is further given by the filled circles.            
See text for more details. 
 }
 \label{fig_eigFreq}
\end{figure}

We are now ready to examine somehow more systematically
    the influence of density asymmetry 
    on the damping rates of resonantly damped quasi-kink modes.
Fixing $[\rhoi/\rhoL, l/d, k_y d, k_z d]$ at $[10, 0.5, 0.5, \pi/50]$,    
    Figure~\ref{fig_eigFreq} presents, by the solid curves, 
    the $\rhoi/\rhoR$-dependencies of
    (a) the oscillation frequency $\omega$,
    (b) the ratio of the damping rate to oscillation frequency $-\gamma/\omega$,        
    and (c) the damping rate $-\gamma$ itself. 
The asterisks in Figures~\ref{fig_eigFreq}a and~\ref{fig_eigFreq}b 
    represent the TB expectations from 
    the approximate expressions in the 
    ``Fully Asymmetric'' and ``Fully Symmetric'' limits
    (Equations~\eqref{eq_TB_FullSym_omgR}
             and \eqref{eq_TB_FullSym_omgI} 
             as well as \eqref{eq_TB_FullAsym_omgR}
             and \eqref{eq_TB_FullAsym_omgI}).
Note that these analytical results are presented largely for reference, 
    given that Figure~\ref{fig_cpTB} has already shown that they 
	are rather close to the resistive results. 
Now examine Figures~\ref{fig_eigFreq}a, where        
    the dash-dotted curve represents the $\rhoi/\rhoR$-dependence
    of the upper bound of the right \Alf\ continuum ($k_z \vaR$).
One sees that the oscillation frequency $\omega$ tends to increase monotonically
    with $\rhoi/\rhoR$, which is intuitively understandable because the effective
    inertia of the system tends to diminish as
    the MHD fluid in the right exterior becomes increasingly rarefied.
Nonetheless, $\omega$ possesses only a rather weak dependence on $\rhoi/\rhoR$,
    being readily overtaken by $k_z \vaR = k_z \vai \sqrt{\rhoi/\rhoR}$
    at some critical $(\rhoi/\rhoR)_{\rm crit}\approx 2.04$. 
Evidently, this $(\rhoi/\rhoR)_{\rm crit}$ is where the right resonance
    starts to be relevant. 
It therefore comes as no surprise that this $(\rhoi/\rhoR)_{\rm crit}$ is 
    reflected in Figure~\ref{fig_eigFreq}b, where the sudden 
    onset of the right resonance somehow leads to a break
    in the $-\gamma/\omega$ curve.     
Regardless, more important to note is that $-\gamma/\omega$ possesses
    an overall nonmonotic dependence on $\rhoi/\rhoR$, attaining some local minimum
    at $(\rhoi/\rhoR)_{\rm min} \approx 3.09$. 
It is interesting to see that $(\rhoi/\rhoR)_{\rm min}$ does not coincide with
    $(\rhoi/\rhoR)_{\rm crit}$, namely the onset of the right resonance
    does not immediately enhance the gross damping efficiency. 
Intuitively speaking, one expects that the relative importance of
    the two resonances is responsible for both
    the overall nonmonotonic $\rhoi/\rhoR$-dependence of $-\gamma/\omega$
    and the difference of $(\rhoi/\rhoR)_{\rm min}$ from $(\rhoi/\rhoR)_{\rm crit}$.
In principle, this intuitive expectation can be readily examined 
    given that the contribution of an individual resonance is measurable by
    $\hat{F}/2\hat{E}$ (see Equation~\eqref{eq_enerCons_1D}), 
    and one can readily perceive how  
    $\hat{F}_{\rm R}$ compares with $\hat{F}_{\rm L}$
    (see the discussions on Figures~\ref{fig_eigFun}a and \ref{fig_eigFun}c).
In practice, however, there arises some difficulty to quantify $\hat{F}$ for a resonance
    due to the need to pinpoint the pertinent DL
    \citep[see e.g.,][for details]{2021ApJ...908..230C}. 
We tackle this by following the empirical approach therein, 
	performing two computations with $\Rm$ being $10^{6}$ and $2\times 10^{6}$, 
    scaling the eigenfunctions in the same way as in Figure~\ref{fig_eigFun},
    and eventually deeming a DL to be
    where $|\tilde{v}_x|$ differs by a factor $\ge 2\times 10^{-4}$
    between the two resistive solutions.
Evaluating $\hat{F}_{\rm L}/2\hat{E}$ and $\hat{F}_{\rm R}/2\hat{E}$ 
    with Equation~\eqref{eq_def_EF}, Figure~\ref{fig_eigFreq}c then
    plots their corresponding values by the open triangles and squares, respectively.
Their sum
    is further presented by the filled circles. 
Two features then follow.  
Firstly, the indirectly evaluated damping rates,
    $(\hat{F}_{\rm L}+\hat{F}_{\rm R})/2\hat{E}$, 
	agree remarkably well with those directly output from the code ($-\gamma$), as evidenced by that the filled circles are threaded by the solid curve.
This further corroborates the remarkable accuracy of the resistive computations.  
Secondly, the contribution to the gross damping rate from the left resonance
    ($\hat{F}_{\rm L}/2\hat{E}$) decreases monotonically with $\rhoi/\rhoR$,
    thereby making it natural to see the decrease of $-\gamma/\omega$
    when $\rhoi/\rhoR$ varies from unity to $(\rhoi/\rhoR)_{\rm crit}$.  
The right contribution, on the other hand, increases monotonically 
    when $\rhoi/\rhoR$ increases from $(\rhoi/\rhoR)_{\rm crit}$,
    tending to the left contribution when
    the ``Fully Symmetric'' configuration is approached. 
After setting in, the right contribution nonetheless only partially
    offsets the reduction in the left contribution, and hence
    the difference between
    $(\rhoi/\rhoR)_{\rm min}$ and $(\rhoi/\rhoR)_{\rm crit}$. 

\begin{figure}
\centering
 \includegraphics[width=.7\columnwidth]{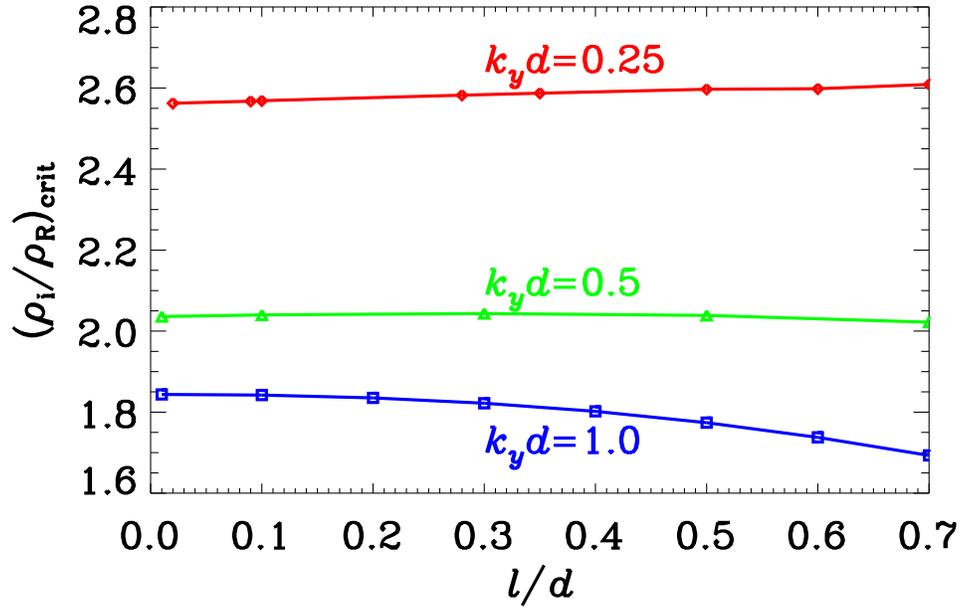}
 \caption{
Dependencies on the dimensionless TL width ($l/d$)
    of the critical density contrast $(\rhoi/\rhoR)_{\rm crit}$, 
    only above which is the right \Alf\ resonance relevant.   
The combination $[\rhoi/\rhoL, k_z d]$ is fixed at $[10, \pi/50]$,
    whereas a number of $k_y d$ are examined as labeled. 
}
 \label{fig_rhocrit}
\end{figure}

Guided by Equation~\eqref{eq_omgFormal}, one may take Figure~\ref{fig_eigFreq}a
    as an approach for locating some specific $(\rhoi/\rhoR)_{\rm crit}$
    for a given combination $[\rhoi/\rhoL, l/d, k_y d, k_z d]$.  
Figure~\ref{fig_rhocrit} capitalizes on this approach to show
    $(\rhoi/\rhoR)_{\rm crit}$ as a function of $l/d$ for a number
    of values of $k_y d$ when $[\rhoi/\rhoL, k_z d]$ is fixed at $[10, \pi/50]$.
In essence, any curve is a dividing line that separates
    the $\rhoi/\rhoR-l/d$ plane into two portions, the right resonance 
    being absent (present) in the portion below (above).
One sees that $(\rhoi/\rhoR)_{\rm crit}$ for a given $k_y d$ 
    possesses only a rather weak dependence on the dimensionless TL width ($l/d$), 
    a behavior that evidently derives from the $l/d$-insensitivity
    of the oscillation frequency $\omega$ (see Figure~\ref{fig_cpTB}a). 
For a given~$l/d$, on the other hand, $(\rhoi/\rhoR)_{\rm crit}$ is seen to 
    possess some stronger dependence on $k_y d$.
Evidently, this dependence follows from the fact that $\omega$ tends
    to decrease somehow appreciably when $k_y d$ increases
    in the examined range.          
These details aside, Figure~\ref{fig_rhocrit} means that
    it makes sense to estimate the critical density contrast $(\rhoi/\rhoR)_{\rm crit}$
    by using the $l=0$ version of Equation~\eqref{eq_TB_DR}, 
    provided that the $l/d$-insensitivity of~$\omega$
    proves sufficiently general. 
This $l=0$ version can be readily implemented by 
    retaining only the first two terms on the LHS of Equation~\eqref{eq_TB_DR},
    writing specifically
\begin{equation}
\label{eq_TB_DRl0}
 \left(
      \dfrac{\kappaL^2}{\mL}\dfrac{\kappaR^2}{m_{\rm R}}
     +\dfrac{\kappai^4}{\mi^2}
 \right)
+\left(\dfrac{\kappaL^2}{\mL}+\dfrac{\kappaR^2}{\mR}\right)
   \dfrac{\kappai^2}{\mi}
   \coth(2\mi d)
=0. 
\end{equation}   
Note that $k_y$ remains involved via the terms 
    $\mi$, $\mL$, and $\mR$ (see Equation~\eqref{eq_def_KappaM}). 
Note further that estimating $(\rhoi/\rhoR)_{\rm crit}$
    with Equation~\eqref{eq_TB_DRl0} is less time-consuming than
    the full resistive eigenmode approach.
However, this does not mean that our resistive approach
    is not worth pursuing, for the $l/d$-insensitivity of $\omega$
    is not known beforehand
    even for the parameters examined in Figure~\ref{fig_rhocrit}.
On top of that, the resistive approach is actually easier to implement
    than the TB formulation (Equation~\eqref{eq_TB_DR})
    when one is interested in, say, the damping rates.  

\subsection{Discussion}  
\label{sec_ResIRA_subDisc}
The observational implications of our results on resonant absorption may be illustrated
    by seeing our equilibrium configuration as a straightened version
    of the much-studied curved arcade system \citep[e.g.,][]{2006A&A...446.1139V,2017A&A...608A.108T}.
Let our $x-z$ plane be identified as the plane of sky (PoS) 
    for the ease of description. 
Likewise, let our configuration be bounded in the axial direction
    by two photospheres at $z=0$ and $z=L$, with $L$ being the arcade length.     
The left (right) exterior then actually corresponds to the outer (inner) ambient corona
    that overlies (underlies) the arcade given that $\rhoL \le \rhoR \le \rhoi$.
Suppose that ideal, zero-beta MHD applies, and that 
    line-tied boundary conditions hold at the bounding planes
    (i.e., $v_{x}=v_{y}=0$).
Suppose further that this system, when initiated with a small-amplitude 
    perturbation in $v_{x}$
    with suitable spatial dependence, evolves into a state where only
    axial standing modes ($k_z = n\pi/L$, with $n=1, 2, \cdots$) 
    are retained and are associated with some     
    specific out-of-plane wavenumber $k_y \gg k_z$.
The following expectations can then be made.
To start, the perturbations in the arcade attenuate, the reason being
    not associated with wave leakage in the $x$-direction but due to
    the energy transfer to \Alfnic\ motions in the TL(s).
Consequently, the~$v_{y}$ perturbations in the TL(s) will feature both 
    a steady growth in magnitude and the development of increasingly fine
    scales in the $x$-direction.
Note that this deduction is made by drawing analogy with    
    the well known behavior of kink oscillations in straight cylinders 
    (see e.g., \citealt{2002ApJ...577..475R,2015ApJ...803...43S};
    also the review by \citetalias{2011SSRv..158..289G}).
Note further that the velocity shear $\partial v_y/\partial x$ may readily
    render some portion of the TL unstable with respect
    to the Kelvin-Helmholtz instability (KHi)  and 
    make visible the KHi-induced vortices
    (see e.g., \citealt{1983A&A...117..220H,1984A&A...131..283B} for
         motivating theories;
     see e.g., \citealt{2008ApJ...687L.115T,2014ApJ...787L..22A,2019FrP.....7...85A}
         for 3D numerical simulations). 
Two regimes may therefore arise in view of
     our Figure~\ref{fig_rhocrit}.
Both the outer and inner edges of the arcade
     may be deformed considerably or
     even become corrugated, if the outer ambient
     is not too different from the inner one. 
However, if the ambient coronae are quite different between the two sides,
     then only the outer edge will show this deformation/corrugation. 
When imaged, the deformation is expected at the apex (two legs) of the arcade 
     when the pertinent vertically polarized kink oscillation is 
          an axial fundamental (the first axial harmonic).
One is therefore allowed to sense, albeit only qualitatively,
     how significantly
     the outer ambient differs from the inner one by looking for
     the morphological differences between the outer and inner edges. 
Evidently, whether this ``morphological seismology'' is feasible needs to be tested
     by imaging observations with high spatial resolution.
Our point, however, is that this does enrich the SAS toolkit because 
     it proves possible to identify 
     a vertically polarized kink mode by using imaging observations alone
          (see \citealt{2008A&A...489.1307W} for TRACE data;
          and  \citealt{2015ApJ...804L..19J} for SDO/AIA results).

\section{Results II: Phase and Group Diagrams}
\label{sec_ResIIgrp}
This section examines how density asymmetry affects the phase and group diagrams
    of oblique quasi-kink modes, in view of the key role that these diagrams
    play in the evolution of a system when locally perturbed
    (see the textbooks by e.g., 
    \citealt{1974Whitham}, hereafter \citetalias{1974Whitham};  
    and \citealt{2019CUP_goedbloed_keppens_poedts}
    for general discussions).
We start by recalling the definition of the 2D wavevector
   $\vec{k} = k_y\uvec{y}+k_z\uvec{z}$, which now needs to be alternatively
   represented by $k$ and $\theta$
   with $k$ being the magnitude ($k=|\vec{k}|$) 
   and~$\theta$ the angle that $\vec{k}$ makes with
   the equilibrium magnetic field $\vec{B}_0$.
The phase velocity is defined as $\vec{v}_{\rm ph} = (\omega/k)\uvec{k}$ 
   with $\uvec{k}$ being the unit vector along $\vec{k}$,
   while the group velocity follows the definition
   $\vec{v}_{\rm gr} = \vgy \uvec{y}+\vgz\uvec{z}$ with
   $\vgy = \partial\omega/\partial k_y$ and 
   $\vgz = \partial\omega/\partial k_z$.  
It is then natural to see only $k_y$ and $k_z$ as independent variables
   in Equation~\eqref{eq_omgFormal}.
Fixing $[\rhoi/\rhoL, l/d]$ at $[10, 0.3]$,
   we find that it suffices to consider two values of $\rhoi/\rhoR$,
   one being $5$ and the other being $1.2$, as far as
   some key influence of density asymmetry on the group diagrams is concerned.
One may question our choice of a non-vanishing $l/d$, given that 
   only the real part ($\omega$) of the eigenfrequency $\Omega$ is involved,
   and that $\omega$ tends to depend on $l/d$ only weakly in the parameter range
   we explore.    
It turns out that the primary results in this section indeed remain almost
   the same if one adopts Equation~\eqref{eq_TB_DRl0} from the outset.
However, choosing a finite $l/d$ makes this section conform better with 
   what we have practiced so far. 
More importantly, it helps avoid the unnecessary impression that
   the results in this section apply only to piece-wise constant
   density profiles.

\subsection{Analytical Expectations in the $l = 0$ Limit}
\label{sec_ResIIgrp_subAnalyt}
This subsection again considers the ``Fully Asymmetric''
   and ``Fully Symmetric'' 
   configurations, but now trying to make some analytical progress on
   the behavior of the phase ($\vec{v}_{\rm ph}$)
        and group velocities ($\vec{v}_{\rm gr}$). 
The reason for us to do this is that the group diagrams 
   tend to be qualitatively different
   in our computations with the two different values of $\rhoi/\rhoR$. 
Suppose that the approximate expressions 
   for the ``Fully Asymmetric'' (``Fully Symmetric'') configuration 
   can somehow reflect what happens when
   $\rhoi/\rhoR = 1.2$ ($\rhoi/\rhoR = 5$). 
The computed group diagrams may then be at least partially understood, which
   we deem necessary because little can be directly inferred from 
   the DR in the TB limit (Equation~\eqref{eq_TB_DR}), let alone the full set of governing equations
   for the resistive EVP. 
It suffices to consider Equation~\eqref{eq_TB_DRl0}, the $l=0$ version
   of Equation~\eqref{eq_TB_DR}. 
Evidently, the oscillation frequencies ($\omega$) for the configurations of interest
   remain largely expressible by Equations~\eqref{eq_TB_FullSym_omgR} and 
   \eqref{eq_TB_FullAsym_omgR} when $\mi^2\approx\mL^2\approx k_y^2$
   or equivalently when $\theta\to 90^\circ$.
It is just that some higher-order corrections may be necessary, given that 
   $\vec{v}_{\rm gr}$ involves not $\omega$ itself but its partial derivatives. 
Furthermore, the limit $\theta\to 0$ is necessary to examine 
   as well.

Consider a ``Fully Symmetric'' configuration ($\rhoR=\rhoL$).
The relevant properties regarding $\vec{v}_{\rm ph}$ and $\vec{v}_{\rm gr}$
   can be summarized as follows.
\begin{itemize}
\item 
When $\theta$ approaches $90^\circ$, 
   the phase speed $\vph = \omega/k$ approaches zero from above
   whereas $\vgy$ does so from below for a given $k$.
In addition, $\vgz$ decreases monotonically with $k$ for a given $\theta$ that
   is sufficiently close to $90^\circ$. 
\item 
When $\theta \to 0$, the $y$-component of the group
   velocity $\vgy\to 0$ for all $k$. 
For a sufficiently small $\theta$, there exists some critical $k$ across which 
   $\vgy$ reverses its sign from negative to positive when $k$ increases. 
\end{itemize}   
We note that the properties for $\theta \to 90^\circ$ 
    follow from Equation~\eqref{eq_TB_FullSym_omgR} in a rather straightforward manner.
Those for $\theta \to 0$, however, require quite some algebra by examining 
    the $l=0$ version of the relevant DR (Equation~\eqref{eq_TB_FullSymDR}) itself. 
We choose to leave out the lengthy derivations, remarking instead that these
    properties have been numerically verified. 
    
Now move on to a ``Fully Asymmetric'' configuration ($\rhoR=\rhoi$). 
Some analytical properties of $\vec{v}_{\rm ph}$ and $\vec{v}_{\rm gr}$
   can be summarized as follows. 
\begin{itemize}
\item When $\theta$ approaches $90^\circ$, 
   the phase speed $\vph = \omega/k$ approaches zero from above for a given $k$,
   and so does $\vgy$ even though $\vgy$ differs little from zero for large $\theta$.
Furthermore, both $\vgy$ and $\vgz$ are essentially
   independent of $k$ for a given $\theta$ that
   is sufficiently close to $90^\circ$.
\item When $\theta\to 0$, the phase speed $\vph$ approaches $\vai$ from above. 
   In addition, $\vgy$ approaches zero for all $k$ and is consistently positive 
      for any sufficiently small $\theta$.           
\end{itemize}
We note that the properties for $\theta\to 90^\circ$ largely follow from
   Equation~\eqref{eq_TB_FullAsym_omgR}, 
   some higher-order corrections being nonetheless necessary.
The net result is that
\begin{equation}
\label{eq_l0_FullAsym_omg_SmTheta}
  \omega 
\approx 
  k_z \ckL 
  \left[
       1-\dfrac{k_z^2}{4k_y^2}
         \left(\dfrac{1-\rhoL/\rhoi}{1+\rhoL/\rhoi}\right)^2
  \right],
\end{equation}
   where the $k_y$- and $k_z$-dependencies can be readily translated into
   the dependencies on $k$ and $\theta$. 
The properties for $\theta \to 0$, on the other hand, are deduced with the approximate
   solution to the $l=0$ version of the relevant DR (Equation~\eqref{eq_TB_FullAsym}) 
   under the assumption $k_y^2 \ll k_z^2$.
This solution writes
\begin{equation}
\label{eq_l0_FullAsym_omg_LgTheta}
  \omega
\approx
  k_z \vai 
  \left[1+\dfrac{k_y^2}{2k_z^2}
         -\dfrac{1}{2(1-\rhoL/\rhoi)}
          \dfrac{k_y^4}{k_z^4}
  \right],    
\end{equation}
   from which the $k$- and $\theta$-dependencies of $\vec{v}_{\rm gr}$ readily follow.

The reason for us to retain the last term in the square parentheses in    
   Equation~\eqref{eq_l0_FullAsym_omg_LgTheta} is connected to the 
   capability for a ``Fully Asymmetric'' configuration
   to guide kink modes with small $\theta$.
Evidently, this capability is measured by $\mR$, 
   which equals $\mi$ in this particular
   case and writes (see Equations~\eqref{eq_def_KappaM} and~\eqref{eq_SolTB_vx})
\begin{equation}
\label{eq_l0_FullSym_mR}
   \mR 
\approx 
   \dfrac{k_z}{\sqrt{1-\rhoL/\rhoi}} 
   \left(\dfrac{k_y^2}{k_z^2}\right).
\end{equation} 
Suppose somehow arbitrarily that only
   those kink modes with $\mR d \ge 1/5$ are observationally relevant,
   where $d$ needs to be understood as some reference spatial scale
   rather than the slab half-width given the absence of the right boundary.
Regardless, suppose further that $kd/\sqrt{1-\rhoL/\rhoi} = 1$.
Equation~\eqref{eq_l0_FullSym_mR} then indicates 
   that the criterion $\mR d \ge 1/5$ translates into 
   $\theta \gtrsim 25^\circ$, an estimate that is substantial enough
   to make the assumption $k_y^2 \ll k_z^2$ questionable.
Then does it still make sense to examine the analytical behavior 
   of $\vec{v}_{\rm ph}$ and $\vec{v}_{\rm gr}$ 
   in this small $\theta$ limit?
The answer is that such an examination remains helpful for understanding
   some key behavior of quasi-kink modes that satisfy, say,
   $\mR d \ge 1/5$ in our $\rhoi/\rhoR=1.2$ computation.    

\subsection{Effects of Density Asymmetry}
\label{sec_ResIIgrp_subDenAsym}
This subsection gathers our numerical results on the 
   phase ($\vec{v}_{\rm ph})$ and group velocities ($\vec{v}_{\rm ph})$ 
   of oblique quasi-kink modes in asymmetric slabs. 
Recall that these results are obtained with the resistive eigenmode approach. 
Recall further that the combination $[\rhoi/\rhoL, l/d]$ is fixed at $[10, 0.3]$. 
Furthermore, we consistently take the condition
   $\mR d \ge 1/5$ as the nominal criterion
   for quasi-kink modes to be of observational relevance.  

\begin{figure}
\centering
 \includegraphics[width=.99\columnwidth]{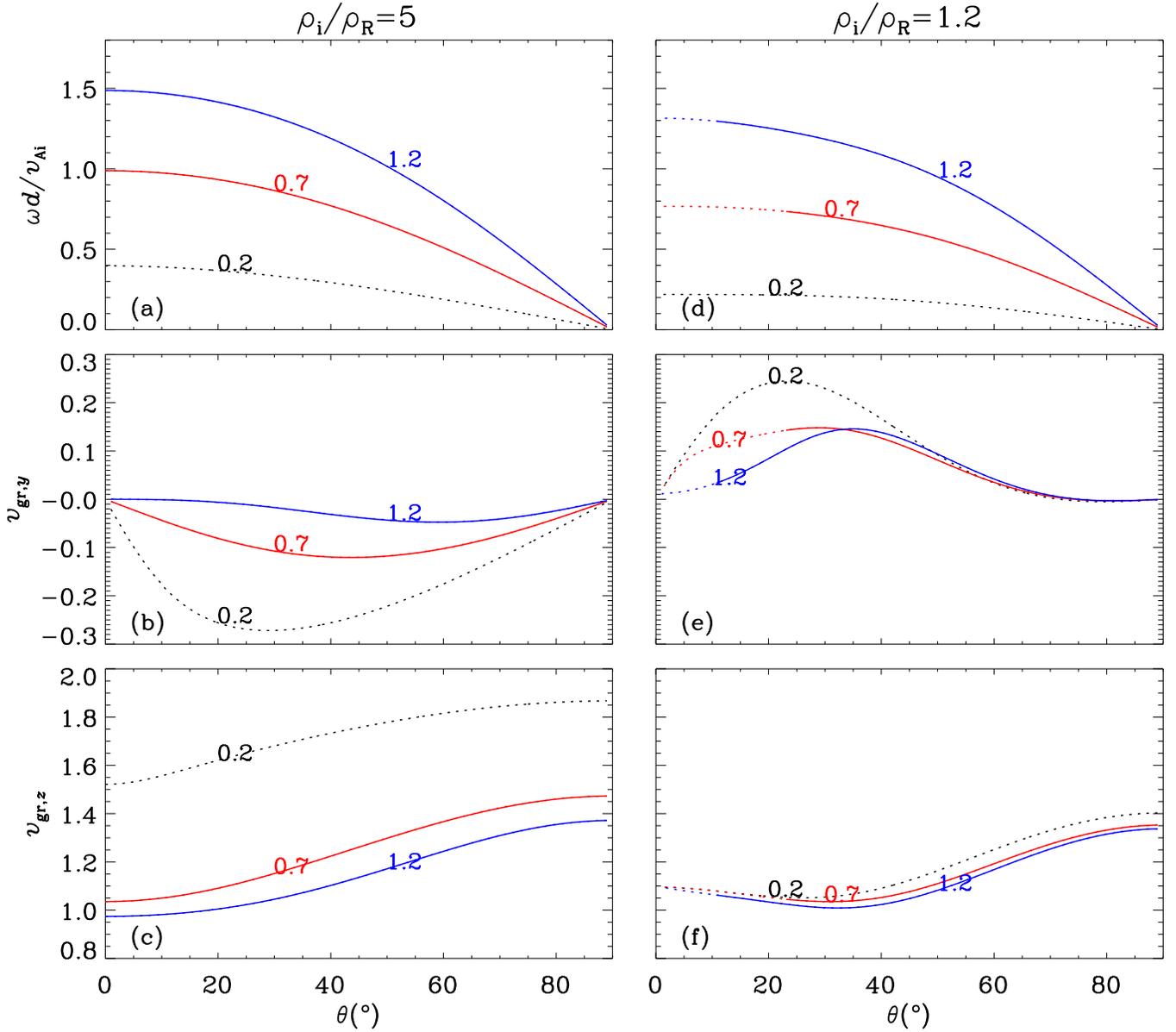}
 \caption{
Dispersion properties of oblique quasi-kink modes in 
	an asymmetric slab configuration, 
	with the obliqueness measured by the angle $\theta$ between 
    the equilibrium magnetic field $\vec{B}_0$
    and a 2D wavevector $\vec{k}=k_y \uvec{y} + k_z \uvec{z}$. 
The combination $[\rhoi/\rhoL, l/d]$ is fixed at $[10, 0.3]$,
    whereas two values of $\rhoi/\rhoR$ are discriminated, one being
    $5$ (the left column) and the other being $1.2$ (right).    
Plotted are the $\theta$-dependencies of
    the oscillation frequency ($\omega$, the top row),
    the $y$-component of the group velocity ($\vgy$, middle),
    and (c) the $z$-component ($\vgz$, bottom).
Several values of $kd$ are examined as labeled. 
The dotted (solid) portion in any curve corresponds to where
    $\mR d <1/5$ ($\mR d  \ge 1/5$), with $\mR d$ measuring
    the capability for the slab to trap oblique quasi-kink modes.
See text for more details.                
}
 \label{fig_vpvgVStht}
\end{figure}

Figure~\ref{fig_vpvgVStht} presents the $\theta$-dependencies
   of the oscillation frequency ($\omega$, the top row),
      the $y$-component of the group velocity ($\vgy$, middle),
  and the $z$-component ($\vgz$, bottom)
  for a number of values of $kd$ as labeled. 
Two values are examined for $\rhoi/\rhoR$, 
   one being $5$ (the left column)
   and the other being $1.2$ (right).
For any quantity, a solid curve is employed 
   to connect its values for those $\theta$ where $\mR d \ge 1/5$, whereas
   a dotted curve is adopted where the opposite is true.    
Note that a curve may not contain any solid or dotted portion,
   with the case for $kd=0.2$ ($kd=0.7$) in the left column
   being an example for the former (latter).
Consider the left column first.
One sees that the numerical results for both large and small values of $\theta$
   are in excellent agreement
   with the analytical properties summarized for a ``Fully Symmetric'' configuration.
This agreement occurs despite that the value adopted for $\rhoi/\rhoR$ in 
   the resistive computations is quite some distance away from $\rhoi/\rhoL$.    
Particularly noteworthy is that $\vgy$ approaches zero from below
   when $\theta\to 90^\circ$ as anticipated (see Figure~\ref{fig_vpvgVStht}b).
Likewise, $\vgy \to 0$ when $\theta\to 0$, 
   and $\vgy$ for a given small $\theta$ is indeed negative (positive)
   when $kd$ is below (above) some threshold.
This threshold $kd$ is nonetheless only marginally smaller than $1.2$, making
   the positive values of $\vgy$ at small $\theta$ for $kd=1.2$ differ little from zero.
For the examined range of $kd$, it then holds in general that 
   $\vgy$ tends to decrease with $\theta$ from zero to some local minimum
   before increasing towards zero when $\theta$ further increases.
It also holds in general that $\omega$ monotonically decreases with $\theta$ for
   a fixed $k$, but is a monotonically increasing function of $k$ when $\theta$
   is fixed (Figure~\ref{fig_vpvgVStht}a). 
Likewise, $\vgz$ turns out to increase (decrease) monotonically with $\theta$ ($k$)
   (Figure~\ref{fig_vpvgVStht}c). 
Now move on to the right column.
The dispersion behavior is substantially more complicated, 
   by which we mean particularly that 
   some analytical expectations summarized for a ``Fully Asymmetric''
   configuration do not apply.
Take the behavior for $\theta \to 90^\circ$ for example.
The numerically computed $\vgy$ is seen to approach zero from below
   rather than from above (Figure~\ref{fig_vpvgVStht}e), 
   and $\vgz$ somehow decreases with $k$ rather than being
   $k$-independent (Figure~\ref{fig_vpvgVStht}f).
These subtleties notwithstanding, our analytical expectations 
   manage to capture some key features for us to proceed, 
   the most noteworthy one being that $\vgy$ starts from being zero when $\theta \to 0$
   and is consistently positive for small $\theta$. 
This feature, together with the analyitcal expectation that $\vgy$ is essentially 
   zero for large $\theta$, then largely explain the behavior 
   for $\vgy$ to be overall positive for the entire range of $\theta$. 
While only three values of $kd$ are presented, a parametric study indicates that   
   the dispersion features, the sign of $\vgy$ in particular, 
   are typical of what happens when $kd$ varies between $0.2$ and $1.2$.
Consequently, that $\mR d < 1/5$ in some range
   of $\theta$ for some $kd$ does not seriously undermine the significance
   of Figure~\ref{fig_vpvgVStht}.          
    
\begin{figure}
\centering
 \includegraphics[width=.99\columnwidth]{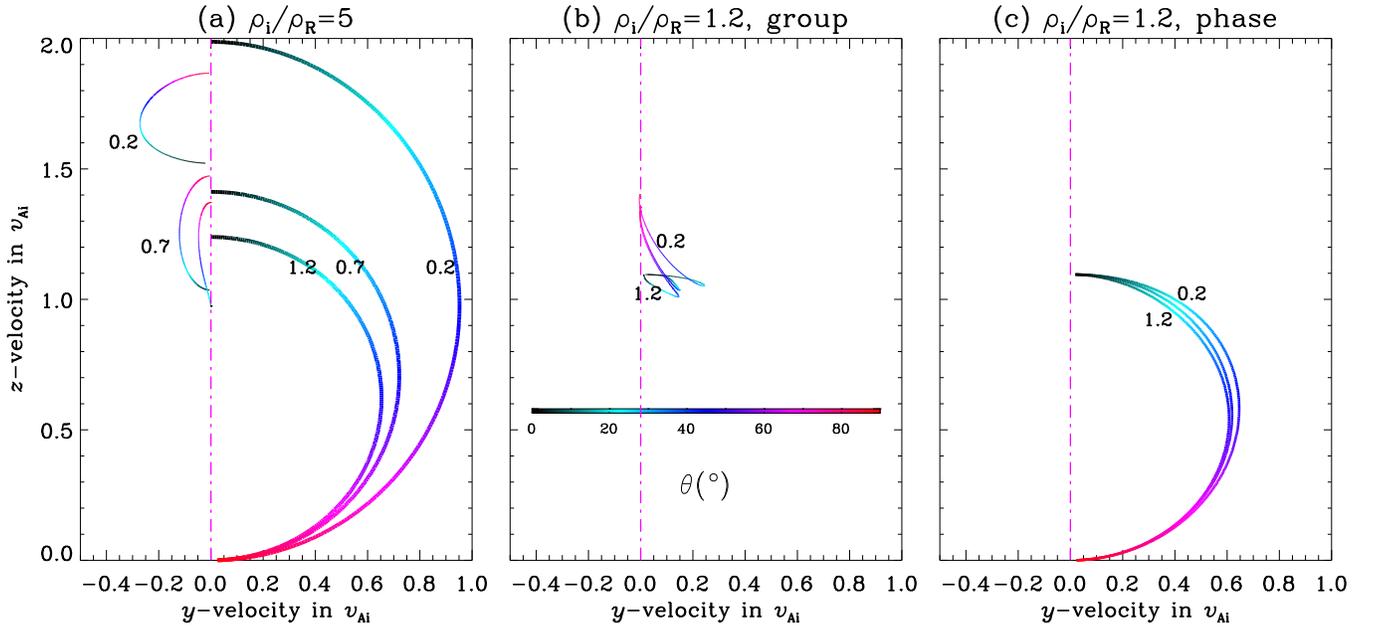}
 \caption{
Phase (the thick curves) and group (thin) diagrams
    of oblique quasi-kink modes in 
	an asymmetric slab configuration.
The combination $[\rhoi/\rhoL, l/d]$ is fixed at $[10, 0.3]$,
    whereas two values of $\rhoi/\rhoR$ are discriminated, one being
    $5$ (the left panel) and the other being $1.2$ (the middle and right panels). 
The vertical dash-dotted line points in the direction
    of the equilibrium magnetic field $\vec{B}_0$.       
Any curve is color-coded by $\theta$, the angle that $\vec{k}$
    makes with $\vec{B}_0$. 
See text for more details.     
}
 \label{fig_pgDiagram}
\end{figure}

Figure~\ref{fig_pgDiagram} further collects      
   the numerical results to produce the relevant phase and group diagrams,
   namely the trajectories that $\vec{v}_{\rm ph}$ (the thick curves)
   and $\vec{v}_{\rm gr}$ (thin) traverse in the velocity plane
   when $\theta$ varies.
The magenta dash-dotted lines represent the vertical axis in this plane,
   pointing in the direction of the equilibrium magnetic field $\vec{B}_0$. 
Any curve is color-coded by $\theta$, and    
   the pertinent value of $kd$ is placed adjacent to a curve when necessary. 
Note that the phase and group diagrams for $\rhoi/\rhoR=5$ 
   are condensed into Figure~\ref{fig_pgDiagram}a,
   whereas those for $\rhoi/\rhoR=1.2$ need to be plotted separately
   to avoid overlapping (Figures~\ref{fig_pgDiagram}b and \ref{fig_pgDiagram}c).
In comparison with Figure~\ref{fig_vpvgVStht}, one sees
   that Figure~\ref{fig_pgDiagram} better visualizes the differences
   in the dispersion behavior when different values are adopted for $\rhoi/\rhoR$.
For instance, the $\vec{v}_{\rm ph}$ trajectories for $\rhoi/\rhoR=5$ 
   are seen to possess a considerably
   stronger $k$-dependence than for $\rhoi/\rhoR=1.2$.    
Any $\vec{v}_{\rm gr}$ trajectory for $\rhoi/\rhoR=1.2$, on the other hand,
   tends to show a more complicated pattern, bending rather abruptly
   (e.g., $kd=1.2$) or even intersecting itself (e.g., $kd=0.7$)
   where the trajectory deviates the most from the $\vec{B}_0$-direction.       
More importantly, the $\vec{v}_{\rm gr}$ and $\vec{v}_{\rm ph}$
   trajectories for $\rhoi/\rhoR=5$
   are seen to lie essentially on opposite sides with respect to $\vec{B}_0$
   (Figure~\ref{fig_pgDiagram}a),
   whereas they are essentially located on the same side when $\rhoi/\rhoR=1.2$
   (see Figures~\ref{fig_pgDiagram}b and \ref{fig_pgDiagram}c).
Evidently, this behavior derives from the difference in the signs 
   of $\vgy$ between the computations with the two different values of $\rhoi/\rhoR$. 
Our discussions on $\mR d$ therefore apply here as well, 
   namely the significance of Figure~\ref{fig_pgDiagram} is not substantially compromised
   by the fact that $\mR d$ may be small in some range of $\theta$ for some
   values of $kd$.

\subsection{Discussion}   
\label{sec_ResIIgrp_subDisc}
We illustrate the observational implications of Figure~\ref{fig_pgDiagram}
    by considering the response of the asymmetric slab system therein
    to a small-amplitude exciter localized around the origin in all three directions.
Recall that $\rhoi/\rhoL=10$. 
For the ease of description, let us suppose that the TLs are absent ($l=0$) even though
    our illustration is expected to hold unless the TLs are excessively thick.  
Suppose further that 
    the initial perturbation is implemented via~$v_x$. 
A close analogy can be drawn with the cylindrical study by
    \citeauthor{2014ApJ...789...48O} 
    (\citeyear{2014ApJ...789...48O}, \citetalias{2014ApJ...789...48O}; also \citealt{2015ApJ...806...56O,2022ApJ...928...33L}).    
In general, both trapped modes 
   (or equivalently ``proper eigenmodes'') 
    and improper continuum eigenmodes are excited.    
However, only trapped modes survive at large times, meaning that 
\begin{equation}
\label{eq_grpIVP_formalSol}
  v_x(x,y,z,t) 
= \int_{-\infty}^{\infty} dk_y 
  \int_{-\infty}^{\infty} dk_z
     \sum_{j} \mathcal{F}_j(x; k_y, k_z) \Exp{i(\omega_j t -k_y y -k_z z)}.   
\end{equation}
Briefly put, Equation~\eqref{eq_grpIVP_formalSol} means that all values
    of $k_y$ and $k_z$ are involved given the localization of the initial perturbation.
An EVP then ensues for any pair $[k_y, k_z]$, the associated DR being
    Equation~\eqref{eq_TB_DRl0}.
The summation in Equation~\eqref{eq_grpIVP_formalSol} incorporates
    all possible eigensolutions ($\omega_j$), 
    with the contribution of the $j$-th solution (namely $\mathcal{F}_j$) 
    determined by both its eigenfunction and the initial perturbation
    (see \citetalias{2014ApJ...789...48O} for technical details).
We proceed by assuming that only transverse fundamental quasi-kink modes 
    are primarily excited, which is not that bold an assumption given the diversity
    of initial perturbations. 
The most straightforward application of Figure~\ref{fig_pgDiagram} then concerns 
    the wave propagation in the $y$- and $z$-directions, meaning that
    it suffices to consider, say, the $x=0$ plane.
Furthermore, we focus on those $(y,z,t)$ where the  
    method of stationary phase (MSP) applies
    (e.g., Chapter~11 in \citetalias{1974Whitham}).
Seeing some $(y,z, t)$ as given, the MSP dictates that     
    $v_x(x=0, y, z, t)$ is dominated by those wavepackets
    with central wavevectors $\vec{K}_n=K_{n,y}\uvec{y}+K_{n,z}\uvec{z}$ that solve
\begin{equation}
\label{eq_MSP_yzt}
\vgy(\vec{K}_n) = y/t, \quad 
\vgz(\vec{K}_n) = z/t.
\end{equation}    
If $(y,z)$ is seen as variable, then 
    the most prominent wave pattern at some given large time can be written as
    (see Equation~(11.41) in \citetalias{1974Whitham})
\begin{equation}
\label{eq_SolMSP}
     v_x(x=0, y, z, t) 
\sim t^{-1}
     \sum_{n}\mathcal{G}_n(K_{n,y}, K_{n,z}) 
         \Exp{i\left[\omega(K_{n,y}, K_{n,z}) t - K_{n,y} y -K_{n,z} z\right]}.   
\end{equation}  

Equation~\eqref{eq_SolMSP} can be seen as a predictive tool,
    the quantitative application of which is nonetheless not straightforward.
One reason for us to say this is that Equation~\eqref{eq_SolMSP},
    while much simpler than
    Equation~\eqref{eq_grpIVP_formalSol},
    still necessitates the summation
    over $n$ because the solution to Equation~\eqref{eq_MSP_yzt} is not unique
    even if only transverse fundamental quasi-kink modes arise.  
Regardless, the following morphological features can be predicted
    for those portions of a large-time wave pattern where most contributions
    come from the wavepackets with central wavevectors that lie in
    the range examined in Figure~\ref{fig_pgDiagram}.
For both values of $\rhoi/\rhoR$ therein,
    these portions will be concentrated around $\vec{B}_0$ (namely the $z$-axis)
	because so are the group trajectories.
We deem this feature somehow striking because the fluid in
    the $x=0$ plane itself is uniform and the waves nonetheless belong to
    the fast family. 
Some subtle differences are then expected in the wave patterns for different
    values of $\rhoi/\rhoR$, to illustrate which point it suffices to 
    consider positive $y$ and $z$.
When $\rhoi/\rhoR=5$, the difference in the signs between $\vgy$ and $\vpy$
    means that some iso-phase curves (say, where $v_x(x=0, y, z, t) =0$)
    will propagate toward the $z$-axis as time proceeds. 
In contrast, that $\vgy$ and $\vpy$ essentially possess the same sign 
    for $\rhoi/\rhoR=1.2$ dictates that the associated iso-phase curves
    propagate away from the $z$-axis.       
Evidently, all these qualitative predictions need to be
    tested against time-dependent 3D numerical simulations, which have yet to conducted
    even for ``Fully Symmetric'' or ``Fully Asymmetric'' slabs to our knowledge
    despite the abundance of 2D ones \citep[e.g.,][]{1993SoPh..144..101M,2006SoPh..236..273O,2013A&A...560A..97P,2021MNRAS.505.3505K,2022MNRAS.515.4055G}.
However, it is safe to conclude that our study highlights the importance of 
    the phase and group trajectories as at least the first step toward
    a thorough understanding of the necessarily complicated 3D wave patterns
    in structured media.
Conversely, these 3D patterns can be employed for seismological purposes, 
    which looks promising given that stereoscopic techniques
    tend to mature with time \citep[see e.g.,][for a review]{2011LRSP....8....5A}
    and have been employed in initial examinations on impulsively excited waves
    in, say, streamer stalks \citep{2020ApJ...893...78D}.

\section{Summary}
\label{sec_conc}

This study was largely motivated by some intensive recent interest in
    small-amplitude magnetoacoustic waves in static, straight, field-aligned, 
    one-dimensional equilibria where the exteriors of a magnetic slab 
    are different between the two sides
    \citep[e.g.,][]{2017SoPh..292...35A,2018ApJ...853..136K,2021ApJ...906..122Z}.
We chose to work with zero-beta MHD such that the inhomogeneity is entirely
    in the equilibrium density $\rho_0(x)$, from which 
    a uniform slab (with density $\rhoi$) and its two uniform exteriors
    (with densities $\rhoL$ and $\rhoR$) are identified. 
By ``left'' we refer to the side that ensures $\rhoL \le \rhoR$. 
Two aspects make our study new, one being that
    $\rho_0(x)$ is not piece-wise constant but varies continuously over some
    transition layer (TL) between the slab and either exterior, 
    the other being that out-of-plane propagation is addressed ($k_y \ne 0$).
Oblique quasi-kink modes, the focus of this study, are therefore absorbed via
    the \Alf\ resonance, their dispersion properties consistently computed 
    with a resistive eigenmode approach. 
We additionally made some analytical progress in the thin-boundary (TB) limit,
    deriving a dispersion relation (DR, Equation~\eqref{eq_TB_DR}) for
    generic asymmetric configurations, and extending previous analytical studies
    on ``Fully Symmetric'' ($\rhoR=\rhoL$)
    or ``Fully Asymmetric'' ($\rhoR=\rhoi$) setups. 
Our findings can be summarized as follows. 
    
Two features stand out in our results on resonant absorption. 
Technically, our resistive computations demonstrated that the TB expectations
    may hold well beyond their nominal range of applicability, thereby 
    corroborating similar conclusions drawn for different coronal configurations. 
Physically, we found that the absorption rates may possess a nonmonotonic
    $\rhoi/\rhoR$-dependence when $\rhoi/\rhoR$ varies from the ``Fully Symmetric''
    to the ``Fully Asymmetric'' limit.
An energetics analysis yields that this behavior results from the difference between
    the two \Alf\ continua, which means particularly that the right resonance
    comes into play only when $\rhoi/\rhoR$ exceeds some threshold
    $(\rhoi/\rhoR)_{\rm crit}$.
Given the likely onset of the Kelvin-Helmholtz instability, 
    we argued that two qualitatively different regimes may arise
    in the morphology of a coronal arcade when oscillating in
    a vertically polarized kink mode.
While only one edge may be deformed when $\rhoi/\rhoR < (\rhoi/\rhoR)_{\rm crit}$, 
    both edges may be subject to deformation or even corrugation
    when the opposite is true. 
   
We also examined oblique quasi-kink modes from the perspective of phase and
    group diagrams, an aspect that has not been addressed to our knowledge. 
We restricted ourselves to only two considerably different values of $\rhoi/\rhoR$.
The group diagrams, namely the trajectories that the group velocity
    traverses in the velocity plane, share the similarity that 
    they are concentrated around the equilibrium
    magnetic field $\vec{B}_0=B_0\uvec{z}$.
However, one key difference between the two sets of computations is that 
    the phase and group trajectories lie essentially on the same side
    (different sides) relative to $\vec{B}_0$ when 
    the equilibrium setup is not far from
    a ``Fully Asymmetric'' (``Fully Symmetric'') one.
We placed our findings in the context of impulsively excited quasi-kink waves in 
    slab-like configurations, expecting the following large-time behavior
    in the $y-z$ cut through the slab axis.
Common to both $\rhoi/\rhoR$, 
    the wave patterns are likely to be highly anisotropic, 
    extending only to a limited angular distance from~$\vec{B}_0$.
However, some iso-phase curves may propagate toward (away from) $\vec{B}_0$
    as time proceeds when the equilibrium is close to 
    a ``Fully Symmetric'' (``Fully Asymmetric'') configuration.

\acknowledgments
This research was supported by the 
    National Natural Science Foundation of China
    (41974200, 41904150, and 11761141002).
We gratefully acknowledge ISSI-BJ for supporting the international team
    ``Magnetohydrodynamic wavetrains as a tool for probing the solar corona''.

\bibliographystyle{aasjournal}
\bibliography{seis_generic}

%

\clearpage

\clearpage
\clearpage

\clearpage

\clearpage

\clearpage

\clearpage

\clearpage

\end{document}